\documentclass[english,nofootinbib]{revtex4-1}
\usepackage{lmodern}

\usepackage[utf8]{inputenc}
\usepackage{xcolor}
\usepackage{babel}
\usepackage{mathtools}
\usepackage{enumitem}
\usepackage{mathrsfs}
\usepackage{amsmath}
\usepackage{amssymb}
\usepackage{graphicx}
\usepackage[normalem]{ulem}
\usepackage[unicode=true,pdfusetitle,
bookmarks=true,bookmarksnumbered=true,bookmarksopen=true,bookmarksopenlevel=1,
 breaklinks=false,pdfborder={0 0 1},backref=false,colorlinks=true]
 {hyperref}
\hypersetup{
 pdfborderstyle=,linkcolor=blue,citecolor=blue,urlcolor={}}



\begin{document}

\title{Bubble universes and traversable wormholes}



\author{Jos\'e P. S. Lemos}
\email{joselemos@ist.utl.pt}

\affiliation{Centro Multidisciplinar de Astrof\'isica -
CENTRA, Departamento de F\'isica,
Instituto Superior T\'ecnico - IST, Universidade de Lisboa -
UL, Av. Rovisco
Pais 1, 1049-001 Lisboa, Portugal}
\author{Paulo Luz}
\email{paulo.luz@ist.utl.pt}
\affiliation{Centro Multidisciplinar de Astrof\'isica -
CENTRA, Departamento de F\'isica,
Instituto Superior T\'ecnico - IST, Universidade de Lisboa -
UL, Av. Rovisco
Pais 1, 1049-001 Lisboa, Portugal}
\affiliation{ Departamento de Matem\'atica, ISCTE - Instituto
Universit\'ario de Lisboa, Portugal}

\begin{abstract}

Bubble universes and traversable wormholes in general relativity can
be realized as two sides of the same concept.  To exemplify it, we
find, display, and study in a unified manner a Minkowski-Minkowski
closed universe and a Minkowski-Minkowski traversable wormhole. By
joining two 3-dimensional flat balls along a thin shell two-sphere of
matter, i.e., a spherical domain wall, into a single spacetime one
gets a Minkowski-Minkowski static closed universe, i.e., a bubble
universe.  By joining two 3-dimensional complements of flat balls
along a thin shell two-sphere of matter, i.e., a spherical throat,
into a single spacetime one gets a Minkowski-Minkowski static open
universe which is a traversable wormhole.  Thus, Minkowski-Minkowski
bubble universes and wormholes can be seen as complementary to each
other.  Is is also striking that these two spacetimes, the
Minkowski-Minkowski bubble universe and the Minkowski-Minkowski
traversable wormhole, have resemblances with two well-known static
universes of general relativity.  The Minkowski-Minkowski static
closed universe, i.e., the Minkowski-Minkowski bubble universe,
resembles in many aspects the Einstein universe, i.e., a static closed
spherical universe homogeneously filled with dust matter and with a
cosmological constant.  The Minkowski-Minkowski static open universe,
i.e., the Minkowski-Minkowski traversable wormhole, resembles the
Friedmann static universe, i.e., a static open hyperbolic universe
homogeneously filled with negative energy density dust and with a
negative cosmological, which is a universe with two disjoint branes,
or branches, that can be considered a failed wormhole.  In this light,
the Einstein static closed universe and the Friedmann static open
universe should also be seen as two sides of the same concept,
i.e., they are complementary to each other.  The scheme is completed
by performing a linear stability analysis for the Minkowski-Minkowski
bubble universe and the Minkowski-Minkowski traversable wormhole and
also by comparing it to the stability of the Einstein static universe
and the Friedmann static universe, respectively.  The complementarity
between bubble universes and traversable wormholes, that exists for
these instances of static spacetimes, can be can carried out for
dynamical spacetimes, indicating that such a complementarity is quite
general.  The overall study suggests that bubble universes and
traversable wormholes can be seen as coming out of the same concept,
and thus, if ones exist the others should also exist.

\end{abstract}

\maketitle

\section{Introduction}

\subsection{Minkowski-Minkowski bubble universe and Minkowski-Minkowski
traversable wormhole}

General relativity is an excellent theory to study universe solutions
and wormhole solutions from which bubble universes and traversable
wormholes can arise as complementary to each other.
To see this, one can attempt to find within
the theory, Minkowski-Minkowski bubble universes and
Minkowski-Minkowski traversable wormholes and study their properties.
One picks up a Minkowski spacetime and at constant time cuts a ball in
it, to obtain two spaces, namely, a 3-dimensional ball with a flat
inside, and an infinite extended 3-dimensional flat space with a hole,
which is the complement of the ball. Then one picks up another
Minkowski spacetime and do the same, to get a second ball and a second
infinite extended flat space with a hole.  If one joins the two
3-dimensional balls along a 2-sphere, a shell containing matter, one
obtains a single 3-space that including time makes altogether a static
closed universe.  If one joins the two complements, i.e., the two
infinite extended 3-dimensional flat spaces with a hole in each,
along a 2-sphere, a shell containing matter, one obtains a different
single 3-space that including time makes altogether another universe,
which is a traversable wormhole. Thus, one has a closed universe,
which can be viewed as a bubble universe, and
its complement, an open universe, which is a traversable wormhole.
To implement the idea of a Minkowski-Minkowski closed universe,
i.e., a bubble universe, and a
Minkowski-Minkowski open universe, i.e., a traversable wormhole, one
uses the equations of general relativity together with the appropriate
thin shell formalism \cite{Israel_1966}.  When one has a thin shell in
an ambient spacetime one has the normal vector to the shell as an
important quantity that will allow to determine how the thin shell
curves in that space, i.e., allows to determine the extrinsic
curvature of the shell, which besides the spacetime metric itself, is
one of the quantities that has to match at both sides of the shell.
Indeed,
to find all possible shell solutions in an ambient spacetime one has to
understand the fact that the normal to a shell can have two relative
directions, such that, for static spherically symmetric spacetimes,
the normal to the shell may point towards or away from the center of
the coordinates.
For intance,
in an ambient Minkowski-Schwarzschild spacetime,
more precisely, for a shell with a
Minkowski interior with a center and a Schwarzschild exterior,
usually called a fundamental shell,
if the
normal points to increasing coordinate radius $r$ in the
exterior,  one has a star
shell, i.e., a shell that represents a star.  In the same ambient
Minkowski-Schwarzschild spacetime, if the normal points to decreasing
coordinate radius $r$ one has a tension shell black hole, i.e., a
shell supported by tension that is in the other side of the
Kruskal-Szekeres diagram as was noted by Katz and Lynden-Bell
\cite{Katz_Lynden-Bell_1991}.  This can also be performed in an
ambient Minkowski-Reissner-Nordstr\"om spacetime, yielding, instead of
two fundamental electrically charged shell spacetimes, a bewildering
variety of fourteen fundamental electrically charged shell spacetimes
with different global spacetime structures \cite{lemosluz}.
Here, in place of using an ambient Minkowski-Schwarzschild or an
ambient Minkowski-Reissner-Nordstr\"om we use an ambient
Minkowski-Minkowski spacetime.

One possibility for a Minkowski-Minkowski spacetime is for a shell
with a Minkowski interior with a center, i.e., a fundamental shell,
such that the normal to the shell points towards decreasing radius $r$
in the Minkowski exterior.  One then finds the Minkowski-Minkowski
closed universe, made of two 3-dimensional flat balls, or sheets,
that are joined at some domain wall, i.e., a 2-sphere shell with
matter, to make a Minkowski-Minkowski bubble universe.  Note that for
a shell with a Minkowski interior with a center such that the normal
to the shell points towards increasing radius $r$ in the Minkowski
exterior yields the trivial global Minkowski spacetime with a zero
shell.

There is yet another possibility for a Minkowski-Minkowski spacetime,
different from the fundamental shell.  It comes from an exotic shell,
i.e., a shell attached to a Minkowski open interior, noting that
interior is just a name since it could as well be called exterior.
For a shell with a Minkowski open interior, when the normal to the
shell points towards increasing $r$ in the Minkowski exterior, one
finds the Minkowski-Minkowski open universe, made of two
3-dimensional flat open infinite sheets that are joined at some
2-sphere with matter, to make a Minkowski-Minkowski traversable
wormhole.  Note that for a shell with a Minkowski open interior such
that the normal to the shell points towards decreasing radius $r$ in
the Minkowski exterior yields the trivial global Minkowski spacetime
with a zero shell.

Universes and wormholes are usually envisaged as distinct objects.
The two Minkowski-Minkowski spacetimes demonstrate that they can be
seen as complementary to each other, i.e., they are two sides of the
same concept.  The concept, i.e., a collection of two Minkowski
spacetimes
together, yields on one side a closed universe, i.e.,
a bubble universe,
and on the
other side an open universe which is a traversable
wormhole. Surely, the
collection of two Minkowski spacetimes
can also lead to two separate
Minkowski spacetimes, but this is the trivial case
and needs not be considered.

\subsection{Einstein static closed spherical
universe and Friedmann static open hyperbolic  universe}

There are two paradigmatic static homogeneous universes in general
relativity. There is the static closed spherical
 universe and there is
the static open  hyperbolic universe, with two
separated branes or branches.
To implement the idea of static uniform universes, one uses general
relativity itself, i.e., Einstein equation
modified to include a
cosmological constant.  From
the staticity condition one imposes that neither the geometry nor the
matter depend on time and from the homogeneous condition
one imposes that
the energy-density is a constant in space.

This implementation, that the Universe, in particular a static
universe, could be described within general relativity, was put
forward by Einstein. In devising a way to realize Mach's principle, a
new interaction, namely, a cosmological constant with repulsion
features, was postulated. General relativity with this new
cosmological interaction
is indeed the first modified theory of gravitation.  This 
repulsive cosmological term, that counterbalances the self
gravitational force due to the energy density of the matter supposed
pressureless, was then used to find a unique static solution for the
Universe which was also assumed to be closed, finite, and spheric
\cite{einstein1917}. In the limiting case that the universe would be
spatially flat, the Einstein universe disappeared in a Minkowski empty
universe. The enforcing of Mach's principle in this way proved to be
a dead end as exemplified by the de Sitter universe with no matter and
only a cosmological constant \cite{desitter}, but the static closed
universe of Einstein was of great impact as it indeed started the
concept of universes. For instance, dynamic closed universes within
general relativity, like the Friedmann \cite{friedman1922}
and Lema\^itre  \cite{lemaitre} expanding universes,
came out of Einstein's static one, which in turn,
due to its instability and propensity to grow, continued to be studied
as a progenitor of expanding closed universes
\cite{edd,bonnor1,harr2,gib1,barr,barcelo}.

Remarkably, Friedmann in his second paper on universes and cosmology
proposed, to start with, an open static hyperbolic universe as to
exhaust the possible static pressureless universes
\cite{friedman1924}.
In the process, a cosmological constant, now
with attraction features, was again introduced to counterbalance the
self gravitational repulsive force due to a matter energy density
necessarily negative.  In the limiting case the universe would be
spatially flat, the Friedmann universe disappeared in a Minkowski empty
universe. The Friedmann static universe can be seen as an anti-Einstein
universe and it inaugurated the concept of open universes. Indeed, it
was used by Friedmann in \cite{friedman1924} to continue the analysis
into dynamic open hyperbolic
universes and it was developed by Harrison
\cite{harr} who, among many other universes, also studied its
stability.  Now, the Friedmann static universe, being hyperbolic, has
two branches, or branes, which fail to communicate to each other by an
infinitesimal separation.  This means that it can be considered a
wormhole, actually, a failed wormhole.  It is not
traversable but almost and
it can be thought as an embryo of a wormhole.

The Einstein static closed universe and the Friedmann static open
universe can be seen as complementary, i.e., they are two
sides of the same concept. The concept here is the
constant spatial curvature of the spacetime, one side 
gives positive spatial curvature, i.e., the Einstein universe, the
other side gives negative spatial curvature, i.e., the Friedmann static
universe.
The trivial case here
would also be two zero curvature spacetimes,
i.e., two separate
Minkowski spacetimes,
and needs not be considered.

\subsection{Bubble universes and traversable wormholes}

Bubble universes and traversable wormholes have been proposed as
structures that might arise if appropriate physical conditions are
available.  Indeed, the Universe in its early phases, of which the
inflationary period is an example, filled with scalar and gauge
fields, may have produced domain walls, cosmic strings, and monopoles,
which can still exist as frozen topological remains of the symmetry
breaking phase transition of that early era.  In this connection, a
setting allowed by the prevailing physical conditions of that early
inflationary era or even of an epoch before it, is that bubble
universes might have unfolded within the Universe and also,
conceivably, systems such as traversable wormholes might have
materialized to connect distant parts of the Universe or distinct
universes.  In addition, a possibility also permitted by the laws of
physics, is that bubble universes and traversable wormholes might be
constructed if sufficient technology is available.  General relativity
is an excellent theory to study universe solutions and wormhole
solutions from which bubble universes and traversable wormholes can
emerge as complementary to each other, and so
they can be seen as duals
of each other, leading to a better understanding of both.

A bubble universe, a universe within the Universe, is a complete
solution of the Einstein's equations.  Bubble universes, together with
baby universes, are universes in themselves, somehow attached to our
one.  They made their appearance in the physics of false vacuum decay
within dynamic bubbles \cite{colemandeluccia}. Its interest and uses
within the inflation theory was seen in \cite{guth}.
The idea of bubble universes taking off out from our Universe was
developed in \cite{blau}, general relativistic dynamic bubble universe
solutions with matter were proposed in \cite{ipser}, several possible
universe decays and corresponding expanfing or contracting domain
walls were thoroughly analyzed in \cite{berezin}, interesting
scenarios with bubbles with different gravitational constants were
proposed in \cite{takamizumaeda}, their intrinsic stability has not
been analyzed, see however \cite{gonzalez}, and bubble universe
astrophysical connections to black holes and their formation were
studied in \cite{dengvilenkin,deng}.

A traversable wormhole, joining two otherwise distinct universes
through two mouths and a throat, is also a complete solution of the
Einstein's equations.
A wormhole is a concept with a history of its own that in a sense was
initiated by Einstein in the celebrated Einstein-Rosen bridge
\cite{einsteinrosen}.  The concept had further developments related to
the quantization of the spacetime geometry \cite{misnerwheeler}, and
it was essential to understand the maximal extension of the
Schwarzschild spacetime, now seen as a white hole being converted to a
black hole through a nontraversable wormhole connecting two separated
asymptotically flat spacetimes \cite{kruskal}, which in turn gave rise
to the notion of multiply connected spacetimes \cite{fullerwheeler}.
Wormholes, in particular traversable wormholes, abound in general
relativity and in gravitational theories.
Traversable wormhole solutions were found first in 
\cite{ellis,bronnikov}, were placed in a proper framework
\cite{morristhorne},
were studied as inflating solutions  in 
\cite{roman},
received the status of being the title of a book in
\cite{visserbook},
had their energy conditions
analyzed in the case of dynamic solutions in
\cite{hochbergvisser},
were built from vacuum stress-energy tensors  in \cite{krasnikov},
were embedded in a cosmological constant
setting  in
\cite{llq},
had the energy conditions at the wormhole's throat
generically reassessed in
\cite{zaslavskii},
were analyzed
in relation to their
shadows and quasinormal modes  in
\cite{bronkonpappas},
had collisions between their own mouths studied  in
\cite{flm}, had a stability analysis
performed  in
\cite{shinkaihayward,sushkovzhang,Lemos_Lobo_2008},
see also \cite{visserbook},
and had their
possible connection to astrophysics
being proposed  in
\cite{kirillov,ovgunetal,piotrovich}.

The bubble universe and traversable wormhole solutions just mentioned
are only a few of the vast number of all the many interesting
solutions found in the literature, and it is not our intention to
proceed with an analysis of so vast a camp.  Usually, bubble universes
appear in dynamic contexts, whereas traversable wormholes are
generically realized in static backgrounds, but of course they can
both be static or dynamic.  Bubble universes and traversable wormholes
can appear as duals of each other, a possibility being when quantum
gravity dominated or was still nonnegligible, bubble universes and
traversable wormholes would form alike out of the spacetime foam, and
would stay stable or metastable structures well into the classical
regime.  We stick to general relativity and to the two static
Minkowski-Minkowski spacetimes and the two static homogeneous
universes of Einstein and Friedmann.
With these four spacetimes it is possible to have two levels of
comparison. On a first level of comparison, on the one hand,
one can compare the two
Minkowski-Minkowski spacetimes between themselves by investigating
their
similarities, and on the other hand, one
can also attempt to compare the two static,
homogenewous, presureless spacetimes of general relativity with
cosmological constant between themselves.  On a second level of
comparison, the two Minkowski-Minkowski spacetimes are put face to
face with the two static spacetimes of Einstein and
Friedmann.
Let us be specific.
When performing  the first level comparison between
the two Minkowski-Minkowski spacetimes, i.e., the Minkowski-Minkowski
bubble universe and the Minkowski-Minkowski traversable wormhole, one
should take some steps, namely, one has to reveal in a unified manner
the two possible nontrivial cases in a Minkowski-Minkowski spacetime,
or more concretely,
one has to find the fundamental shell spacetime, which
is a closed bubble universe, and find also the exotic shell spacetime,
which is an open traversable wormhole universe. In doing so, one
classifies and analyzes the possible junctions of Minkowski
spacetimes through a static, timelike, thin matter shell, which are
the two nontrivial cases just mentioned, the trivial one being the no
shell pure Minkowski spacetime. A linear stability study of these
spacetimes completes the comparison.  Through this example, bubble
universes and traversable wormholes can now be understood in a unified
light, in the sense that the Minkowski-Minkowski bubble universe and
the Minkowski-Minkowski traversable wormhole are two sides of the same
concept, in which instance, if one exists it makes a case to the
existence of the other.
When performing the first level comparison between the two static,
homogenewous, presureless spacetimes with cosmological constant, i.e.,
the Einstein closed universe and the Friedmann open universe, one
should take some steps, namely, one has to reveal in a unified manner
these two possible nontrivial cases, and display them in a new light.
A linear stability study of these spacetimes completes the
analysis.  In this new light, the Einstein and Friedmann static
universes can be compared, they are seen anew as a bubble universe and
a failed wormhole, respectively.
On the second level of comparison,  the two Minkowski-Minkowski
spacetimes are put face to face with the two static homogeneous
spacetimes, to find that the Minkowski-Minkowski closed
universe, a bubble universe, goes along with the Einstein closed
universe, which can then be seen then as a bubble universe, and the
Minkowski-Minkowski open universe, a traversable wormhole, goes along
with the Friedmann open universe, which is a failed wormhole.
This comparison shows some striking
resemblances between those spacetimes on several grounds. 
In this sense, the Einstein and Friedmann static universes are really
seen anew as a bubble universe and a failed wormhole, respectively,
and also they are an example that bubble universes and traversable
wormholes can be perceived in a unified light.
The complementarity, or duality, between general relativistic bubble
universes and traversable wormholes exists for these examples of
static spacetimes.  One can carry out this idea for dynamical
spacetimes and show that the complementarity, or duality, considered
here is quite generic.  Moreover, following this rationale, if one
finds inflating bubble universe solutions, one should be able to find
the corresponding inflating wormhole solutions, and vice versa, so
that, for instance, a given solution already found in one of the sides
could help in looking for the complementary solution in the other
side.

The paper is organized as follows.
In Sec.~\ref{sec:closedandopenminkuniverses}, we formalize
in a unified way the two
possible junctions of two identical Minkowski spacetime regions and we
perform a linearized stability analysis of the Minkowski-Minkowski
universes.  We then build in detail the Minkowski-Minkowski static
closed universe, i.e., a bubble universe, and the Minkowski-Minkowski
static open universe, i.e., the Minkowski-Minkowski static traversable
wormhole.
In Sec.~\ref{sec:closedandopenEFuniverses}, we formalize
in a unified way the two
possible static homogeneous universes and we display a linearized
stability analysis of them.  We then display in detail the Einstein
static spherical closed universe and we compare explicitly the
Minkowski-Minkowski static closed universe, i.e., the
Minkowski-Minkowski bubble universe, 
with the Einstein static
spherical closed universe, and display the Friedmann static hyperbolic
open universe, i.e., the failed wormhole, and compare explicitly the
Minkowski-Minkowski static open universe, i.e., the
Minkowski-Minkowski static traversable wormhole, with the Friedmann
failed wormhole.
In Sec.~\ref{Sec:Conclusions} we conclude.
Throughout the paper we work in geometrized units system where the
constant of gravitation $G$ and the speed of light $c$ are set to one,
$G=1$ and $c=1$, and use the metric signature $\left(-+++\right)$.

\section{Minkowski-Minkowski closed universe and Minkowski-Minkowski
open universe}
\label{sec:closedandopenminkuniverses}

\subsection{Minkowski-Minkowski universes: Formal solutions and
stability}
\label{Sec:Junction}

\subsubsection{Solutions}

The
Einstein field equations are
\begin{equation}
G_{\alpha\beta}=
8\pi T_{\alpha\beta}\,.\label{eq:EFE1}
\end{equation}
where $G_{\alpha\beta}=R_{\alpha\beta}-
\frac{1}{2}g_{\alpha\beta}\mathcal{R}$ is the Einstein tensor,
$R_{\alpha\beta}$ and $\mathcal{R}$ are the Ricci tensor and Ricci
scalar, respectively,
$g_{\alpha\beta}$ is the metric,  $T_{\alpha\beta}$ is the
stress-energy tensor, and Greek indices run from o to 3
with 0 representing a time component and 1, 2, and 3
representing space components.
One wants to join consistently
two solutions of Einstein field equations
and the Israel formalism provides the method needed
to make
the junction between two different
general relativistic
spacetime regions~\citep{Israel_1966}. Consider then two
spacetime manifolds with boundary, one is
$\mathcal{M}_{\mathrm{i}}$ with
metric $g_{\mathrm{i}}$ and
the other is $\mathcal{M}_{\mathrm{e}}$
with metric $g_{\mathrm{e}}$. The spacetimes
$(\mathcal{M}_{\mathrm{i}},
g_{\mathrm{i}})
$
and $(\mathcal{M}_{\mathrm{e}},
g_{\mathrm{e}})
$
are solutions of the theory of general relativity
and are to be glued together at a common
boundary, forming a new spacetime $\mathcal{M}$. In brief,
$\mathcal{M}$ is partitioned by
an hypersurface $\mathcal{S}$ into two regions,
the regions $\mathcal{M}_{\mathrm{i}}$
and $\mathcal{M}_{\mathrm{e}}$. The
formalism applies directly to a
hypersurface $\mathcal{S}$ that can 
be either timelike or spacelike, the
extension to the case of a null
boundary hypersurface can also be done
with care.

We assume that it is possible to
formally define a common
coordinate system
$\left\{ x^{\alpha}\right\} $ on both sides of the hypersurface
$\mathcal{S}$. We also
admit
the existence of a
normal vector field $n$, well defined
on both sides of $\mathcal{S}$, which is orthogonal to
the matching hypersurface
at each point. We choose $n$
to point from $\mathcal{M}_{\mathrm{i}}$ to $\mathcal{M}_{\mathrm{e}}$
and, without loss of generality, 
$
n^{\alpha}n_{\alpha}=\varepsilon
$,
where $n^{\alpha}$ are the components of $n$ in the coordinate system
$\left\{ x^{\alpha}\right\} $ and $\varepsilon$ is $+1$ or $-1$
depending on $n$ being spacelike or timelike, respectively. The null
case has $\varepsilon=0$ and it would have to be treated separately
which we do not do here.
For a timelike normal vector field $n$ one has that the
corresponding 
hypersurface $\mathcal{S}$
is spacelike, and 
for a spacelike $n$ one has that the
corresponding 
hypersurface $\mathcal{S}$
is timelike, and vice versa. Then, assuming
$\left\{ y^{a}\right\} $ to represent a local coordinate system on
$\mathcal{S}$, the normal vector field $n$ must be orthogonal at
each point to the tangent vectors to the hypersurface $\mathcal{S}$,
$e_{a}\equiv
\frac{
\partial\;\;}{\partial y^{a}}$, such that 
$
e_{a}^{\alpha}n_{\alpha}=0
$,
with $e_{a}^{\alpha}\equiv
\frac{\partial x^{\alpha}}{\partial y^{a}}$.
The
induced metric on $\mathcal{S}$ as seen
from each region $\mathcal{M}_{\mathrm{i}}$
and $\mathcal{M}_{\mathrm{e}}$, is 
$
h_{\mathrm{i}\,{ab}}
=
g_{\mathrm{i}\,\alpha\beta}e_{a}^{\alpha}e_{b}^{\beta}$,
$
h_{\mathrm{e}\,ab} 
=
g_{\mathrm{e}\,\alpha\beta}e_{a}^{\alpha}e_{b}^{\beta}
$, respectively, 
where $g_{\mathrm{i}\,\alpha\beta}$ and
$g_{\mathrm{e}\,\alpha\beta}$ are the components of
the metrics $g_{\mathrm{i}}$ and
$g_{\mathrm{e}}$ in the coordinate system $\left\{
x^{\alpha}\right\} $.  Notice that, in general,
the induced metric on
$\mathcal{S}$ by each metric $g_{\mathrm{i}}$ and
$g_{\mathrm{e}}$  may not coincide, hence we use the
notation $h_{\mathrm{i}\,ab}$ and
$h_{\mathrm{e}\,ab}$ to refer to the metric induced
by the spacetime structure of $\mathcal{M}_{\mathrm{i}}$ or
$\mathcal{M}_{\mathrm{e}}$, respectively.
The extrinsic curvature $K_{\mathrm{i}\,ab}$ or
$K_{\mathrm{e}\,ab}$ of the hypersurface
$\mathcal{S}$, as an
embedded manifold in $\mathcal{M}_{\mathrm{i}}$ or
$\mathcal{M}_{\mathrm{e}}$, respectively, is defined as
$
K_{\mathrm{i}\,ab} 
= e_{a}^{\alpha}e_{b}^{\beta}
\nabla_{\mathrm{i}\,\alpha}n_{\beta}$,
$K_{\mathrm{e}\,ab} 
= e_{a}^{\alpha}e_{b}^{\beta}
\nabla_{\mathrm{e}\,\alpha}n_{\beta}
$,
where $\nabla_{\mathrm{i}}$ and $\nabla_{\mathrm{e}}$ represent the
covariant derivatives with respect to $g_{\mathrm{i}}$
or $g_{\mathrm{e}}$.
Their traces are
$K_{\mathrm{i}} 
= h_\mathrm{i}^{ab} 
K_\mathrm{i\,ab}$, and
$K_{\mathrm{e}} 
= h_\mathrm{e}^{ab} 
K_\mathrm{i\,ab}$, respectively.

Now, we need to give
the conditions under which the matching of the
two spacetimes $\mathcal{M}_{\mathrm{i}}$ and
$\mathcal{M}_{\mathrm{e}}$
form a valid solution of the Einstein field equations,
Eq.~(\ref{eq:EFE1}).
Following the Israel
formalism, to join the two spacetimes
$\mathcal{M}_{\mathrm{i}}$ and $\mathcal{M}_{\mathrm{e}}$ at
$\mathcal{S}$, such that the union of $g_\mathrm{i}$
and $g_\mathrm{e}$ forms a valid solution to the
Einstein field equations~\eqref{eq:EFE1}, two junction conditions must
be verified at the matching surface $\mathcal{S}$:
(i) 
The induced metric $h_{ab}$
as seen from each region
$\mathcal{M}_{\mathrm{i}}$ and $\mathcal{M}_{\mathrm{e}}$,
must be the same, i.e.,
\begin{equation}
\left[h_{ab}\right]=0\,.\label{eq:1st_junct_cond}
\end{equation}
(ii)
If the extrinsic curvature $K_{ab}$ is not
the same on both sides of the boundary $\mathcal{S}$, then a thin
shell is present at $\mathcal{S}$ with stress-energy tensor $S_{ab}$
given by 
\begin{equation}
-\frac{\varepsilon}{8\pi}\left(\left[K_{ab}\right]-
h_{ab}\left[K\right]\right)=S_{ab}
\,,\label{eq:2nd_junct_cond}
\end{equation}
where
$\left[K_{ab}\right]$
represents the
difference of $K_{ab}$
as seen from each sub-manifold at $\mathcal{S}$,
i.e., 
$\left[K_{ab}\right]\equiv
\left.
{K_{\mathrm{e}\,ab}
}\right|_{\mathcal{S}}-
\left.
{K_{\mathrm{i}\, ab}}\right|_{\mathcal{S}}$,
and similarly for 
$\left[K\right]$,
and we use the notation
${K_{\mathrm{i}\,ab}}
\equiv
K_{ab}\left(\mathcal{M}_{\mathrm{i}}\right)
$
and 
${K_{\mathrm{e}\,ab}}
\equiv
K_{ab}\left(\mathcal{M}_{\mathrm{e}}\right)
$
to refer to $K_{ab}$ defined in $\mathcal{M}_{\mathrm{i}}$
or $\mathcal{M}_{\mathrm{e}}$, respectively, and
 similarly for 
$K$.

The Minkowski spacetime with line element $ds^2=-dt^2 +dr^2+
r^2d\Omega^{2}$, where $t$ and $r$ are the time and radial
coordinates, and $d\Omega^{2}\equiv d\theta^{2}+\sin^{2}\theta
d\varphi^{2}$, with $\theta$ and $\varphi$ being the angular
coordinates, is a solution of Einstein equations, see
Eq.~(\ref{eq:EFE1}), in fact the simplest solution.  We assume now
that the interior and exterior spacetime have Minkowski line elements,
and find and analyze all possible junctions of two Minkowski
spacetimes through a static, thin matter shell.  Using the Israel
formalism, we consider two spacetimes, $\mathcal{M}_{\mathrm{i}}$ and
$\mathcal{M}_{\mathrm{e}}$, each endowed with the Minkowski metric
tensor field glued together at a common hypersurface, $\mathcal{S}$.
To apply the formalism, we will have to find the induced metric and
extrinsic curvature induced on an embedded hypersurface of each
spacetime, $\mathcal{M}_{\mathrm{i}}$ or
$\mathcal{M}_{\mathrm{e}}$.
We will start by making the analysis in
the interior spacetime, $\mathcal{M}_{\mathrm{i}}$, and extend the
results to the exterior spacetime, $\mathcal{M}_{\mathrm{e}}$.

The interior Minkowski spacetime, $\mathcal{M}_{\mathrm{i}}$, is
characterized by the following line element, in spacetime spherical coordinates,
\begin{equation}
ds_{\mathrm{i}}^2=-dt_{\mathrm{i}}^2
+dr_{\mathrm{i}}^2+
r_{\mathrm{i}}^2d\Omega^{2}\,,\label{eq:Mink_metric_interior}
\end{equation}
where $t_{\mathrm{i}}$ and $r_{\mathrm{i}}$ are the time and radial
coordinates, respectively, measured by a free-falling observer in
$\mathcal{M}_{\mathrm{i}}$, and again $d\Omega^{2}\equiv
d\theta^{2}+\sin^{2}\theta d\varphi^{2}$, with $\theta$ and $\varphi$
being the angular coordinates.  On one hand, for the solution itself
we are interested in studying the case where the hypersurface
$\mathcal{S}$ is timelike and static, i.e., static as seen from an
observer free falling in the interior Minkowski spacetime. On the
other hand for the stability analysis that we will
take we have to allow for the
hypersurface to be dynamical, an so we compute the induced
metric and extrinsic curvature of $\mathcal{S}$ allowing for a
dynamical shell and when needed we take the static
solution. In the study of the properties of the matter shell
at $\mathcal{S}$, we will then restrict the setup to the static
case. Since we assume the hypersurface to be timelike, it is
convenient to choose the coordinates on $\mathcal{S}$ to be $\left\{
y^{a}\right\} =\left(\tau,\theta,\varphi\right)$, where $\tau$ is the
proper time measured by an observer comoving with $\mathcal{S}$.  In
this coordinate system, it follows that $e_{\tau}\equiv u$, where $u$
is the 4-velocity of an observer comoving with the shell. The
hypersurface $\mathcal{S}$, as seen from the interior
$\mathcal{M}_{\mathrm{i}}$ spacetime, is parameterized by $\tau$, such
that the surface's radial coordinate is described by a function
$R_{\mathrm{i}}=R_{\mathrm{i}}\left(\tau\right)$.  Then, the
4-velocity $u_\mathrm{i}$, where
the subscript $\mathrm{i}$ is not an index
and as before denotes interior, as seen from the interior
spacetime is given by
$u_\mathrm{i}=
\left(\frac{dt_{\mathrm{i}}}{d\tau},
\dot{R}_{\mathrm{i}},0,0\right)$,
where overdot represents the derivative with respect to the proper
time, i.e., $\dot{R}_{\mathrm{i}}\equiv
\frac{dR_{\mathrm{i}}}{d\tau}$.
Since $\mathcal{S}$ is a timelike hypersurface, it must verify
$u_{\mathrm{i}\,\alpha}
u_\mathrm{i}^\alpha=-1$,
therefore we find
$u_{\mathrm{i}}=
\left(\sqrt{1+\dot{R}_\mathrm{i}^2},
\dot{R}_{\mathrm{i}},0,0\right)$,
where we chose $\frac{dt_{\mathrm{i}}}{d\tau}>0$ as
we assume $u_\mathrm{i}$
to point to the future.
The expression for the 4-velocity of an observer comoving with
$\mathcal{S}$ and the condition
$e_{a}^{\alpha}n_{\alpha}=0$ allow us to find
the components of the normal vector field to $\mathcal{S}$, $n$, as
seen from the interior spacetime $\mathcal{M}_{\mathrm{i}}$,
$n_\mathrm{i}=\xi_{\mathrm{i}}
\left(\dot{R}_{\mathrm{i}},
\sqrt{1+\dot{R}_\mathrm{i}^2},0,0\right)$,
where $\xi_{\mathrm{i}}$ is a normalization factor.
Using $n^{\alpha}n_{\alpha}=\varepsilon$,
Eq.~\eqref{eq:Mink_metric_interior}, and the condition that
the normal
vector field $n_\mathrm{i}$ is spacelike, yields
$\xi_{\mathrm{i}}=\pm1$. Now, defining $\nabla_\mathrm{i}
r_\mathrm{i}$ as
the gradient of $r_{\mathrm{i}}$, the choice $\xi_{\mathrm{i}}=+1$
or $\xi_{\mathrm{i}}=-1$ represents whether the inner product
$g_{\mathrm{i}}\left(n_{\mathrm{i}},
\nabla_\mathrm{i} r_{\mathrm{i}}\right)>0$
or
$g_{\mathrm{i}}\left(n_{\mathrm{i}},
\nabla_\mathrm{i} r_{\mathrm{i}}\right)<0$,
respectively. Under the Israel formalism both
values for $\xi_{\mathrm{i}}$
are possible and we shall consider both cases.
Using the induced metric equation,
$h_{\mathrm{i}\,{ab}}
 =
g_{\mathrm{i}\,\alpha\beta}e_{a}^{\alpha}e_{b}^{\beta}$,
and
$u_{\mathrm{i}}=
\left(\sqrt{1+\dot{R}_\mathrm{i}^2},
\dot{R}_{\mathrm{i}},0,0\right)$,
we find that the induced metric on $\mathcal{S}$ by the spacetime
$\mathcal{M}_{\mathrm{i}}$, is such that
the line element can be written as
$
\left.ds^{2}\right|_{\mathcal{S_{\mathrm{i}}}}=-
d\tau^{2}+R_\mathrm{i}^2d\Omega^{2}
$.
Gathering these results, we can compute the
components of the extrinsic curvature of $\mathcal{S}$ as seen from
$\mathcal{M}_{\mathrm{i}}$,
$K^\mathrm{i}_{ab}$.  In
the case where the matching surface $\mathcal{S}$ is timelike and
spherically symmetric, the non-null components of the extrinsic
curvature are given by, dropping here the superscript i
to not overcrowd the notation, $K_{\tau\tau}=-a^{\alpha}n_{\alpha}$,
$K_{\theta\theta}=\nabla_{\theta}n_{\theta}$,
$K_{\varphi\varphi}=\nabla_{\varphi}n_{\varphi}$, where
$a^{\alpha}\equiv u^{\beta}\nabla_{\beta}u^{\alpha}$
represents the components of the
4-acceleration of an observer comoving with $\mathcal{S}$.
Taking into account
Eq.~\eqref{eq:Mink_metric_interior}
and 
$u_{\mathrm{i}}=
\left(\sqrt{1+\dot{R}_\mathrm{i}^2},
\dot{R}_{\mathrm{i}},0,0\right)$,
we
find that the non-trivial components of the exterior curvature as seen
from the interior Minkowski spacetime are given by
${{K_\mathrm{i}}^\tau}_{\tau}=
\xi_{\mathrm{i}}\frac{\ddot{R}_{\mathrm{i}}}{\sqrt{1+
\dot{R}_\mathrm{i}^2}}$,
and 
${{K_\mathrm{i}}^\theta}_{\theta}=
{{K_\mathrm{i}}^\varphi}_{\varphi}=
\xi_\mathrm{i}
\frac{\sqrt{1+
{\dot{R}}_\mathrm{i}^2}}{R_{\mathrm{i}}}$,
where the induced metric $h_\mathrm{i}^{ab}$
associated with the hypersurface line element was
used to raise the indices.
Since we assume the shell to be static, one
has
$\dot{R}_{\mathrm{i}}=0$: So in this
static case, in brief,
one has, that
the 
4-velocity $u_\mathrm{i}$ and the normal 
$n_\mathrm{i}$ are
$
u_{\mathrm{i}}=
\left(1,
0,0,0\right)$
and
$
n_\mathrm{i}=\xi_{\mathrm{i}}
\left(0,
1,0,0\right)
$,
the line element on the shell is
\begin{equation}
\left.ds^{2}\right|_{\mathcal{S_{\mathrm{i}}}}=-
d\tau^{2}+R_\mathrm{i}^2d\Omega^{2}\,,
\label{eq:induced_metric_Interior}
\end{equation}
and the extrinsic curvature is given by
\begin{equation}
{{K_\mathrm{i}}^\tau}_{\tau}=0\,,\quad\quad
{{K_\mathrm{i}}^\theta}_{\theta}=
{{K_\mathrm{i}}^\varphi}_{\varphi}=
\frac{\xi_\mathrm{i}}{R_{\mathrm{i}}}
\,.
\label{eq:Extrinsic_curvature_Interior}
\end{equation}
Under the Israel formalism both values for $\xi_\mathrm{i}$
are possible and we shall consider both cases.

The exterior Minkowski spacetime, $\mathcal{M}_{\mathrm{e}}$, is
characterized by the following line element, in spacetime spherical coordinates,
\begin{equation}
ds_{\mathrm{e}}^2=-dt_{\mathrm{e}}^2
+dr_{\mathrm{e}}^2+
r_{\mathrm{e}}^2d\Omega^{2}\,,\label{eq:Mink_metric_exterior}
\end{equation}
where $t_{\mathrm{e}}$ and $r_{\mathrm{e}}$ are the time and radial
coordinates, respectively, measured by a free-falling observer in
$\mathcal{M}_{\mathrm{e}}$, and again $d\Omega^{2}\equiv
d\theta^{2}+\sin^{2}\theta d\varphi^{2}$, with $\theta$ and $\varphi$
being the angular coordinates. Since the setup is the same as the one
for the interior we will sketch the calculations briefly in order to
be complete. 
For a timelike hypersurface it is
convenient to choose the coordinates on $\mathcal{S}$ to be $\left\{
y^{a}\right\} =\left(\tau,\theta,\varphi\right)$, where $\tau$ is the
proper time measured by an observer comoving with $\mathcal{S}$.  In
this coordinate system, it follows that
the 4-velocity $u$ of an observer comoving with the shell
is given by $e_{\tau}\equiv u$.
The
hypersurface $\mathcal{S}$, as seen from the exterior
$\mathcal{M}_{\mathrm{e}}$ spacetime, is parameterized by $\tau$, such
that the surface's radial coordinate is described by a function
$R_{\mathrm{e}}=R_{\mathrm{e}}\left(\tau\right)$.  
Then, $u_\mathrm{e}=
\left(\frac{dt_{\mathrm{e}}}{d\tau},
\dot{R}_{\mathrm{e}},0,0\right)$,
where
the subscript $\mathrm{e}$ is not an index
and as before denotes exterior, and 
$\dot{R}_{\mathrm{e}}\equiv
\frac{dR_{\mathrm{e}}}{d\tau}$.
Since $\mathcal{S}$ is timelike, one has
$u_{\mathrm{e}\,\alpha}
u_\mathrm{e}^\alpha=-1$,
so
$u_{\mathrm{e}}=
\left(\sqrt{1+\dot{R}_\mathrm{e}^2},
\dot{R}_{\mathrm{e}},0,0\right)$,
where we chose $\frac{dt_{\mathrm{e}}}{d\tau}>0$ as
we assume $u_\mathrm{e}$
to point to the future.
From $e_{a}^{\alpha}n_{\alpha}=0$, 
one finds 
the components of the normal vector field to $\mathcal{S}$, $n$, as
seen from the exterior spacetime, namely
$n_\mathrm{e}=\xi_{\mathrm{e}}
\left(\dot{R}_{\mathrm{e}},
\sqrt{1+\dot{R}_\mathrm{e}^2},0,0\right)$,
where $\xi_{\mathrm{e}}$ is a normalization factor.
Using $n^{\alpha}n_{\alpha}=1$
and Eq.~\eqref{eq:Mink_metric_exterior}, yields
$\xi_{\mathrm{e}}=\pm1$.
Defining $\nabla_\mathrm{e}
r_\mathrm{e}$ as
the gradient of $r_{\mathrm{e}}$, the choice $\xi_{\mathrm{e}}=+1$
or $\xi_{\mathrm{e}}=-1$ represents whether the inner product
$g_{\mathrm{e}}\left(n_{\mathrm{e}},
\nabla_\mathrm{e} r_{\mathrm{e}}\right)>0$
or
$g_{\mathrm{e}}\left(n_{\mathrm{e}},
\nabla_\mathrm{e} r_{\mathrm{e}}\right)<0$,
respectively.
Usingthe induced metric
equation,
$h_{\mathrm{e}\,ab} 
=
g_{\mathrm{e}\,\alpha\beta}e_{a}^{\alpha}e_{b}^{\beta}$,
and
$u_{\mathrm{e}}=
\left(\sqrt{1+\dot{R}_\mathrm{e}^2},
\dot{R}_{\mathrm{e}},0,0\right)$,
we find that the induced metric on $\mathcal{S}$ by the spacetime
$\mathcal{M}_{\mathrm{e}}$, is such that
the line element can be written as
$
\left.ds^{2}\right|_{\mathcal{S_{\mathrm{e}}}}=-
d\tau^{2}+R_\mathrm{e}^2d\Omega^{2}
$.
The non-null components of the extrinsic
curvature are here given by, dropping here the superscript e
to not overcrowd the notation, $K_{\tau\tau}=-a^{\alpha}n_{\alpha}$,
$K_{\theta\theta}=\nabla_{\theta}n_{\theta}$,
$K_{\varphi\varphi}=\nabla_{\varphi}n_{\varphi}$, where
$a^{\alpha}\equiv u^{\beta}\nabla_{\beta}u^{\alpha}$
represents the components of the
4-acceleration of an observer comoving with $\mathcal{S}$.
Taking into account
Eq.~\eqref{eq:Mink_metric_exterior}
and $u_{\mathrm{e}}=
\left(\sqrt{1+\dot{R}_\mathrm{e}^2},
\dot{R}_{\mathrm{e}},0,0\right)$
the non-trivial components of the exterior curvature as seen
from the exterior Minkowski spacetime are given by
${{K_\mathrm{e}}^\tau}_{\tau}=
\xi_{\mathrm{e}}\frac{\ddot{R}_{\mathrm{e}}}{\sqrt{1+
\dot{R}_\mathrm{e}^2}}$,
and 
${{K_\mathrm{e}}^\theta}_{\theta}=
{{K_\mathrm{e}}^\varphi}_{\varphi}=
\xi_\mathrm{e}
\frac{\sqrt{1+
{\dot{R}}_\mathrm{e}^2}}{R_{\mathrm{e}}}$,
where the induced metric $h_\mathrm{e}^{ab}$
associated with the hypersurface line element was
used to raise the indices.
Since we assume the shell to be static, one
has
$\dot{R}_{\mathrm{e}}=0$. So in this case
one has that
the 
4-velocity $u_\mathrm{e}$ and the normal 
$n_\mathrm{e}$ are
$
u_{\mathrm{e}}=
\left(1,
0,0,0\right)$
and 
$n_\mathrm{e}=\xi_{\mathrm{e}}
\left(0,
1,0,0\right)
$,
the line element on the shell is
\begin{equation}
\left.ds^{2}\right|_{\mathcal{S_{\mathrm{e}}}}=-
d\tau^{2}+R_\mathrm{e}^2d\Omega^{2}\,,
\label{eq:induced_metric_Exterior}
\end{equation}
and the extrinsic curvature is given by
\begin{equation}
{{K_\mathrm{e}}^\tau}_{\tau}=0\,,\quad\quad
{{K_\mathrm{e}}^\theta}_{\theta}=
{{K_\mathrm{e}}^\varphi}_{\varphi}=
\frac{\xi_\mathrm{e}}{R_{\mathrm{e}}}
\,.
\label{eq:Extrinsic_curvature_Exterior}
\end{equation}
Both
values for $\xi_{\mathrm{e}}$
are possible and we shall consider both cases.

To complete the solution we have find the properties of the matter at
the thin shell.
Indeed, having found previously the necessary expressions 
at the hypersurface $\mathcal{S}$ as seen from the
$\mathcal{M}_{\mathrm{i}}$ and $\mathcal{M}_{\mathrm{e}}$ spacetimes,
we can now use the Israel formalism to glue
together the two spacetimes.  Imposing the first junction condition,
Eq.~\eqref{eq:1st_junct_cond}, and using
Eqs.~\eqref{eq:induced_metric_Interior} 
and
\eqref{eq:induced_metric_Exterior} for the induced metrics,
we find that the radial coordinate
of $\mathcal{S}$ as seen from the interior and exterior spacetimes,
$R_{\mathrm{i}}$ and $R_{\mathrm{e}}$, respectively, must be the same,
$R_{\mathrm{i}}=R_{\mathrm{e}}$. We then denote by $R$ the value of
the radial coordinate of $\mathcal{S}$ as seen from both spacetimes,
\begin{equation}
R\equiv
R_{\mathrm{i}}=
R_{\mathrm{e}}
\,.
\label{eq:R}
\end{equation}
We also assume that the stress-energy tensor $S_{ab}$ of
the thin shell on $\mathcal{S}$, can be cast in a perfect fluid form
$
S_{ab}=\sigma u_{a}u_{b}+p\left(h_{ab}+u_{a}u_{b}\right)
$,
where $\sigma$ is the energy per unit area, $p$ is the tangential
pressure of the fluid, $h_{ab}$ is the induced metric on $\mathcal{S}$,
and $u_a$ is the fluid's 3-velocity on $\mathcal{S}$.
Using the appropriate equations we find
$
S^{\tau}{}_{\tau}=-\sigma
$,
and
$
S^{\theta}{}_{\theta}=S^{\varphi}{}_{\varphi}=p
$.
Having found previously the expressions for the
extrinsic curvature of the hypersurface $\mathcal{S}$ as seen from the
$\mathcal{M}_{\mathrm{i}}$ and $\mathcal{M}_{\mathrm{e}}$ spacetimes,
and knowing that the stress-energy tensor of the shell is that of a
perfect fluid, we are can now use
the second junction condition, Eq.~\eqref{eq:2nd_junct_cond}.
Applying it to
our spherically symmetric problem gives that the only nontrivial
components of the stress-energy tensor $S^{a}{}_{b}$ are given by
$S^{\tau}{}_{\tau}=\frac{1}{4\pi}\left[K^{\theta}{}_{\theta}\right]$,
and
$S^{\theta}{}_{\theta}=S^{\varphi}{}_{\varphi}=
\frac{1}{8\pi}\left[K^{\tau}{}_{\tau}\right]+\frac{1}{4\pi}
\left[K^{\theta}{}_{\theta}\right]$.
Using then for the interior,
${{K_\mathrm{i}}^\tau}_{\tau}=
\xi_{\mathrm{i}}\frac{\ddot{R}}{\sqrt{1+
\dot{R}^2}}$,
${{K_\mathrm{i}}^\theta}_{\theta}=
{{K_\mathrm{i}}^\varphi}_{\varphi}=
\xi_\mathrm{i}
\frac{\sqrt{1+
{\dot{R}}^2}}{R}$,
and for the exterior,
${{K_\mathrm{e}}^\tau}_{\tau}=
\xi_{\mathrm{e}}\frac{\ddot{R}}{\sqrt{1+
\dot{R}^2}}$,
and 
${{K_\mathrm{e}}^\theta}_{\theta}=
{{K_\mathrm{e}}^\varphi}_{\varphi}=
\xi_\mathrm{e}
\frac{\sqrt{1+
{\dot{R}}^2}}{R}$,
we find
$\sigma  =
\left(\xi_{\mathrm{i}}-\xi_{\mathrm{e}}\right)
\frac{\sqrt{1+\dot{R}^{2}}}{4\pi R}$
and
$p  =-
\left(\xi_{\mathrm{i}}-\xi_{\mathrm{e}}\right)
\frac{
R\ddot{R}+\dot{R}^{2}+1}
{8\pi R\sqrt{1+\dot{R}^{2}}}$.
From these two equations 
we derive the following conservation law for the shell
$
\dot{\sigma}+\frac{2\dot{R}}{R}\left(\sigma+p\right)=0
$.
For a static shell, the time derivatives 
are zero and so,
using directly if one wishes
Eqs.~\eqref{eq:Extrinsic_curvature_Interior}
and~\eqref{eq:Extrinsic_curvature_Exterior},
one finds
\begin{align}
\sigma & =\quad
\left(\xi_{\mathrm{i}}-\xi_{\mathrm{e}}\right)
\frac{1}{4\pi R}
\label{sig1}\,,\\
p & =-
\left(\xi_{\mathrm{i}}-\xi_{\mathrm{e}}\right)
\frac{1}{8\pi R}
\,.
\label{press1}
\end{align}
From Eqs.~(\ref{sig1})
and~(\ref{press1})
we derive 
\begin{equation}
\sigma+2p=0\,.
\label{17}
\end{equation}
The matter of the thin shell obeys necessarily this equation of state,
namely, $p=-\frac12\sigma$, for a Minkowski-Minkowski static
spacetime.
From Eqs.~(\ref{sig1})
and~(\ref{press1}), we see that
for $\xi_{\mathrm{i}}=\xi_{\mathrm{e}}$,
and so 
$\frac12(\xi_{\rm i}-\xi_{\rm e})=0$
we get the trivial case,
$\sigma=0$ and $p=0$.
For $\xi_{\mathrm{i}}=1$ and $\xi_{\mathrm{e}}=-1$, and so 
$\frac12(\xi_{\rm i}-\xi_{\rm e})=1$,
we get
$\sigma=\frac{1}{2\pi R}$ and $p=-\frac{1}{4\pi R}$.
For
$\xi_{\mathrm{i}}=-1$ and $\xi_{\mathrm{e}}=1$,
and so 
$\frac12(\xi_{\rm i}-\xi_{\rm e})=-1$,
we get
$\sigma=-\frac{1}{2\pi R}$ and $p=\frac{1}{4\pi R}$.
All cases obey Eq.~(\ref{17}),
i.e., the relation between the surface energy density
$\sigma$ and the surface pressure $p$
is independent of $\xi_{\mathrm{i}}$
or $\xi_{\mathrm{e}}$.
Besides the trivial case, i.e.,
the Minkowski universe which has 
$\frac12(\xi_{\rm i}-\xi_{\rm e})=0$,
there are two possible universes at this juncture, the
Minkowski-Minkowski static closed
universe which has $\frac12(\xi_{\rm i}-\xi_{\rm e})=1$,
and 
the
Minkowski-Minkowski static open
universe which has $\frac12(\xi_{\rm i}-\xi_{\rm e})=-1$.

\subsubsection{Linearized stability analysis for Minkowski-Minkowski
universes}

An important question regarding the Minkowski-Minkowski static
universe solutions, i.e., the Minkowski-Minkowski closed and open
universes, is if these are stable under perturbations. Here we will
discuss the linear stability of the Minkowski-Minkowski solutions that
we have found by analyzing the equation of motion of the shell near
the static configuration.

To study the linear stability of the Minkowski-Minkowski static
universe solution we have to find the evolution equation for the shell
radius $R$ and analyze the behavior of these solutions as we perturb
the spacetime.  The analysis can be done in a unified way by making
use of the parameter $\frac12(\xi_{\rm i}-\xi_{\rm e})$.
The equation of motion of the thin shell
previously found, namely,   
$\sigma  =
\left(\xi_{\mathrm{i}}-\xi_{\mathrm{e}}\right)
\frac{\sqrt{1+\dot{R}^{2}}}{4\pi R}$, can be
inverted and put in the form
\begin{equation}
\dot{R}^{2}+V(R)=0\,,\label{eq:stability_EoM_general}
\end{equation}
where
a dot means derivative with respect to time $t$,
and the potential $V(R)$ is given by
\begin{equation}
V(R)=1-\left(\frac{4\pi
R\sigma}{\xi_{\mathrm{i}}-\xi_{\mathrm{e}}}\right)^{2}\,.
\label{eq:stability_potential}
\end{equation}
A thin matter shell is stable 
if and only if the potential $V(R)$ at the shell's position
is at a local
minimum, i.e., if $V'(R)=0$ and
$V''(R)\geq0$, with the equality providing the 
marginal neutral  case and
where a prime denotes the derivative with respect to $R$.
Thus, we have to calculate the matter properties
and its derivatives.
All these properties are functions of the shell radius $R$,
namely, $\sigma=\sigma(R)$,
$p=p(R)$,
$\sigma'=\sigma'(R)$, and $p'=p'(R)$.
To find an expression for $\sigma'(R)$, we
consider
the conservation law for the shell already found,
namely,
$
\dot{\sigma}+\frac{2\dot{R}}{R}\left(\sigma+p\right)=0$, i.e.,
$
\dot{\sigma}=-\frac{2\dot{R}}{R}\left(\sigma+p\right)$. 
Using the inverse function theorem, we have $\sigma'
=\frac{\dot{\sigma}}{\dot{R}}$,
and so
\begin{equation}
\sigma'=-\frac{2\left(\sigma+p\right)}{R}\,.
\label{eq:stability_sigma_prime}
\end{equation}
We have also to analyze the derivatives of
the potential at the static configuration. From
Eq.~\eqref{eq:stability_potential} we get
$V'(R)=-\frac{32\pi^{2}R\sigma}{\left(\xi_{\mathrm{i}}-
\xi_{\mathrm{e}}\right)^{2}}\left(\sigma+R\sigma'\right)$.
Taking another derivative we get
$
V''(R)=\frac{32\pi^{2}}{\left(\xi_{\mathrm{i}}-
\xi_{\mathrm{e}}\right)^{2}}\left[\left(\sigma+R\sigma'\right)
\left(\sigma+2p\right)+R\sigma\left(\sigma'+2p'\right)\right]
$.
Now, if we introduce Eq.~(\ref{eq:stability_sigma_prime}) into
$V'(R)$, we have 
$V'(R)=\frac{32\pi^{2}R\sigma}{\left(\xi_{\mathrm{i}}-
\xi_{\mathrm{e}}\right)^{2}}\left(\sigma+p\right)$.
To analyze $V''(R)$ we
have to find an expression for $p'(R)$.
We 
assume that the thin matter shell is composed of cold  matter
such that it verifies a generic equation of state of the form
$p=p\left(\sigma\right)$. Then, we can define the
parameter $\eta\left(\sigma\right)=\frac{\partial p}{\partial\sigma}$
such that, $p'=\eta\sigma'$. Hence, using
$\sigma'+2p'=\sigma'\left(1+2\eta\right)$ and
Eq.~\eqref{eq:stability_sigma_prime} we can write $V''(R)$
as $
V''(R)=-\frac{32\pi^{2}}{\left(\xi_{\mathrm{i}}-
\xi_{\mathrm{e}}\right)^{2}}\left[2\sigma\left(\sigma+p\right)
\left(1+2\eta\right)+\left(\sigma+2p\right)^{2}\right]
$. In brief, the derivatives of the potential $V(R)$
are
\begin{equation}
V'(R)=\frac{32\pi^{2}R\sigma}{\left(\xi_{\mathrm{i}}-
\xi_{\mathrm{e}}\right)^{2}}\left(\sigma+p\right)\,, 
\label{eq:stability_V_prime}
\end{equation}
and
\begin{equation}
V''(R)=-\frac{32\pi^{2}}{\left(\xi_{\mathrm{i}}-
\xi_{\mathrm{e}}\right)^{2}}\left[2\sigma\left(\sigma+p\right)
\left(1+2\eta\right)+\left(\sigma+2p\right)^{2}\right]\,.
\end{equation}
where $\eta\left(\sigma\right)=
\frac{\partial p}{\partial\sigma}$.

We now linearize the equation of motion for the shell given by 
Eq.~\eqref{eq:stability_EoM_general} around a static solution. Defining
$R_{0}$ as the circumferential radius of the static thin shell and
assuming the potential $V$ to be a differentiable function at $R_{0}$,
we can expand the potential
given in Eq.~\eqref{eq:stability_potential} around
$R_{0}$ as
\begin{equation}
V(R)=V(R_{0})+V'(R_{0})
\left(R-R_{0}\right)+\frac{1}{2}V''(R_{0})
\left(R-R_{0}\right)^{2}\,,
\label{eq:stability_potential_expansion}
\end{equation}
plus higher order terms of
$\mathcal{O}\left[\left(R-R_{0}\right)^{3}\right]$.
A thin matter shell with  radius $R_{0}$ is stable
or neutrally stable
if and only if the potential $V(R)$
satisfies 
$V'(R_{0})=0$ and
$V''(R_{0})\geq0$.
For a shell with radius $R_{0}$
the static
solutions found in the previous section are characterized
generically by
the following 
expressions,
$\sigma  =
\frac{\xi_{\mathrm{i}}-\xi_{\mathrm{e}}}{4\pi R_{0}}$
and
$
p  =-\frac{\xi_{\mathrm{i}}-\xi_{\mathrm{e}}}{8\pi R_{0}}
$,
see Eqs.~(\ref{sig1})
and~(\ref{press1}),
where $\xi_{\mathrm{i}}\neq\xi_{\mathrm{e}}$
for the nontrivial solutions. Substituting
Eq.~\eqref{sig1} into
Eq.~\eqref{eq:stability_potential} we find $V(R_{0})=0$, as
expected.
Substituting Eqs.~\eqref{sig1}
and~\eqref{press1}
into Eq.~\eqref{eq:stability_V_prime} and evaluating
it at the static solution we find $V'(R_{0})=0$. 
Evaluating $V''(R)$ at the static solution, $R=R_{0}$, and
using again Eqs.~\eqref{sig1}  and
\eqref{sig1}, we find
$V''(R_{0})=0$. In brief,
\begin{equation}
V'(R_{0})=0\,,
\end{equation}
and
\begin{equation}
V''(R_{0})=0\,.
\label{stabMM}
\end{equation}
Moreover, all higher order
derivatives of the potential go to zero at $R_{0}$ for the static
solutions.

Gathering these calculations, we conclude that, besides the trivial
case, i.e., the Minkowski universe which has $\frac12(\xi_{\rm
i}-\xi_{\rm e})=0$ and is trivially neutraly stable, there is the
Minkowski-Minkowski static closed universe which has $\frac12(\xi_{\rm
i}-\xi_{\rm e})=1$ and is nontrivially neutraly stable, and the
Minkowski-Minkowski static open universe which has $\frac12(\xi_{\rm
i}-\xi_{\rm e})=-1$ and is also nontrivially neutraly stable.  This
neutral stability means that if we slightly displace the thin shell,
it will simply stay at the new radius. This confirms our expectation,
as the interior and exterior spacetimes are both described by a
Minkowski, i.e., flat, solution.

\subsection{Minkowski-Minkowski universes: Geometry and physics}

\subsubsection{
Minkowski-Minkowski static closed universe: A bubble universe}

Here we display a Minkowski-Minkowski static closed universe as a
solution of general relativity. We rely on the results presented
above.  We assume that the Minkowski line element is valid for a
region, which we call interior $\mathcal{M}_{\mathrm{i}}$, up to a
radius $R$, i.e., $0\leq r_\mathrm{i}\leq R$, where $r_\mathrm{i}$
denotes the interior radial coordinate.  We join this region to
another region, which we call exterior $\mathcal{M}_\mathrm{e}$, where
the exterior radial coordinate is denoted by $r_\mathrm{e}$.  The
junction is done at a common hypersurface $\mathcal{S}$ with
circumferential radius $r_\mathrm{i}=r_\mathrm{e}=R$.  Thus, the whole
spacetime is composed by the two regions plus the common hypersurface,
which is a domain wall, i.e., a
thin shell.  The common hypersurface $\mathcal{S}$ is
assumed to be static.  Assuming the existence of a vector field $n$,
normal, at each point, to the common hypersurface $\mathcal{S}$, we
have found that the solution depends on the orientation of this normal
field $n$.  For each region, $\mathcal{M}_{\mathrm{i}}$ and
$\mathcal{M}_{\mathrm{e}}$, the orientation of the normal is encoded
in a single parameter, namely, one for the interior,
$\xi_{\mathrm{i}}$, and one for the exterior, $\xi_{\mathrm{e}}$. In
both cases, $\xi_{\mathrm{i}}$ and $\xi_{\mathrm{e}}$ can have values
$+1$ or $-1$.  The value $+1$ indicates that the normal points in the
direction of increasing radial coordinate and the value $-1$ indicates
that the normal points in the direction of decreasing radial
coordinate.  The solutions with $\xi_{\mathrm{i}}=\xi_{\mathrm{e}}$
are trivial as the resulting spacetime is simply the full Minkowski
flat universe. Here we consider the first non-trivial solution, i.e.,
$\xi_{\mathrm{i}}=+1$ and $\xi_{\mathrm{e}}=-1$.  This static solution
represents the case where the normal vector field $n$ points in the
direction of increasing radial coordinate as seen from the interior
spacetime $\mathcal{M}_{\mathrm{i}}$ and points in the direction of
decreasing radial coordinate as seen from the exterior spacetime
$\mathcal{M}_{\mathrm{e}}$. Since $n$ is assumed to point from
$\mathcal{M}_{\mathrm{i}}$ to $\mathcal{M}_{\mathrm{e}}$ this implies
that in the exterior region one also has $0\leq r_\mathrm{e}\leq R$.
Thus, this solution is composed by two spatially compact Minkowski
spacetime regions glued together at the common boundary $\mathcal{S}$.  This
solution then represents a Riemann flat spacetime everywhere except at
$\mathcal{S}$. Overall it is a static closed Minkowski-Minkowski
universe, for which the line element can be written as, see also
Eqs.~(\ref{eq:Mink_metric_interior})
and~(\ref{eq:Mink_metric_exterior}),
\begin{equation}\label{eq:Mink_metric}
ds^{2}=-dt^{2}+dr_\mathrm{i}^{2}+r_\mathrm{i}^{2}d\Omega^{2},
\quad0\leq r_\mathrm{i}\leq R\,,
\quad\quad\;\;\;
ds^{2}=-dt^{2}+dr_\mathrm{e}^{2}+r_\mathrm{e}^{2}d\Omega^{2},
\quad 0\leq r_\mathrm{e}\leq R\,.
\end{equation}
Now we turn to the properties of the matter at
the domain wall, or shell, at $\mathcal{S}$.
In this case, putting $\xi_{\mathrm{i}}=+1$ and $\xi_{\mathrm{e}}=-1$
into Eqs.~(\ref{sig1}) and~(\ref{press1}), the energy density $\sigma$
and the tangential pressure $p$ at the domain wall are given by
\begin{equation}
\sigma=\;\;\frac{1}{2\pi R}\,,
\label{eq:sigma_Closed_Universe}
\end{equation}
\begin{equation}
p=-\frac{1}{4\pi R}\,.
\label{eq:pressure_Closed_Universe}
\end{equation}
This solution is then characterized by the presence of a surface layer
in the form of a domain wall at radius $R$, separating two Minkowski
halves. The thin domain wall is composed of a perfect fluid with
positive energy density and is supported by tension such that it
obeys the equation of state $\sigma+2p=0$. Moreover, from
Eqs.~\eqref{eq:sigma_Closed_Universe} and
\eqref{eq:pressure_Closed_Universe}, we find that the following
inequalities are verified: $\sigma\geq0$, $\sigma+p\geq0$,
$\sigma+2p\geq0$ and $\sigma\geq\left|p\right|$, therefore, the matter
at the domain wall verifies the null, weak, strong, and dominant
pointwise energy conditions. Since the effective  mass
$m$ can be defined by the quantity
$\sigma+2p$ and this latter is zero, one has
$m=0$. So the domain wall yields no total mass $m$ as it should to
have Minkowski spacetime on both sides of the domain wall.  The volume
of this universe is $V=\frac{8\pi}{3}R^3$.

\begin{figure}[h]
\centering\includegraphics[height=0.13\paperheight]
{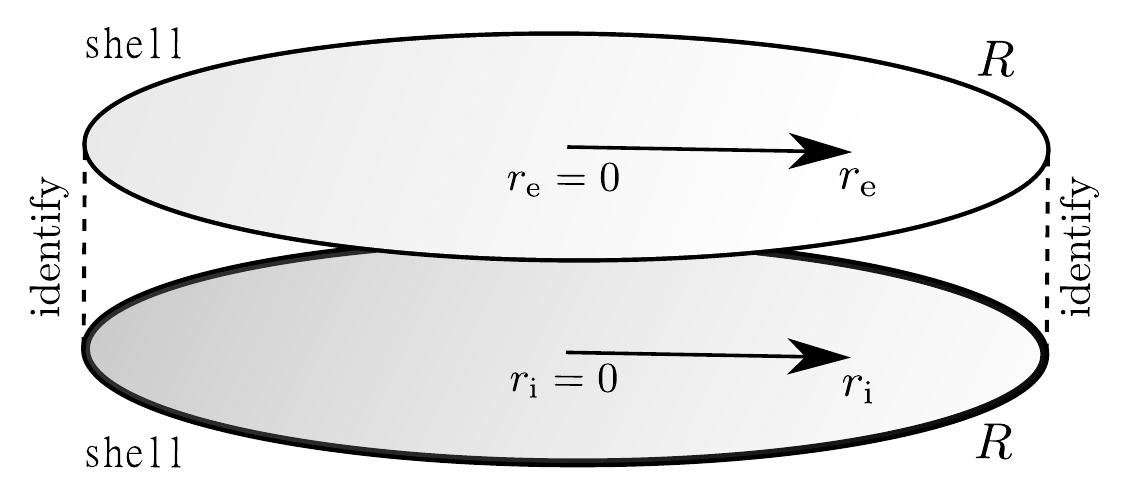}
\caption{\label{fig:time_slice_closed_universe}
Embedding diagram of a $t=\text{constant}$ and $\theta=\frac{\pi}{2}$
slice of the Minkowski-Minkowski static closed universe in
3-dimensional Euclidean space.  The interior coordinate is
in the range $0\leq
r_{\mathrm{i}}\leq R$, the exterior coordinate is
in the range $0\leq
r_{\mathrm{i}}\leq R$, the radius of the domain wall, or
shell, is $R$, and the borders
of the circumferences should be identified.
}
\end{figure}

\begin{figure}[h]
\centering\includegraphics[height=0.25\paperheight]
{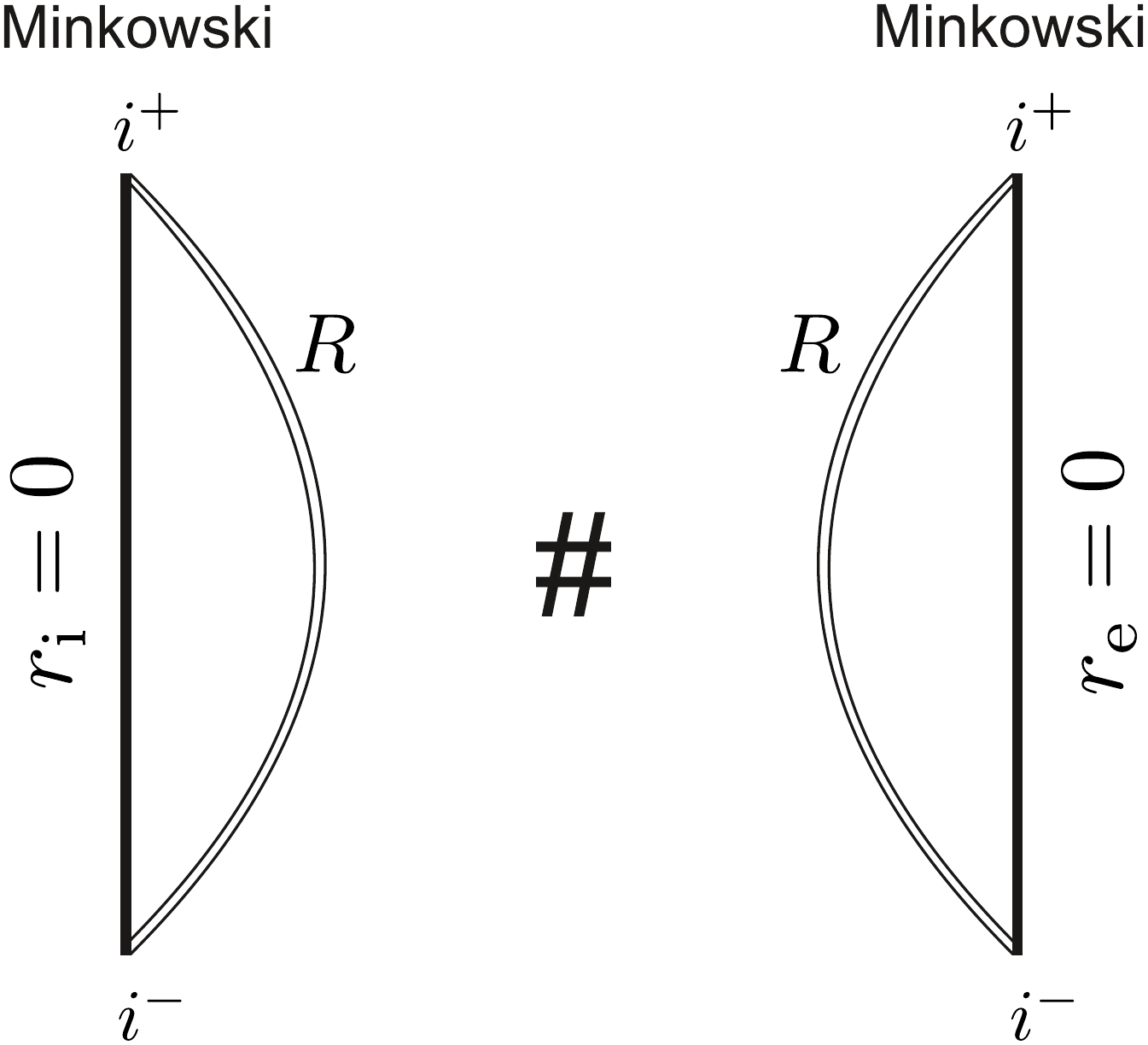}
\caption{\label{fig:Carter-Penrose_closed_universe}
Carter-Penrose diagram of the Minkowski-Minkowski static closed
universe.  The hash symbol \# represents the connected sum of the
spacetime manifold. The symbols $i^-$ and $i^+$ represent past and
future causal
infinity, respectively.
}
\end{figure}

The spatial structure and the causal structure are also important to
analyze.  A time slice $t={\rm constant}$ of the spacetime gives that
the 3-space is a highly squashed 3-sphere, i.e., it is made of two
copies of two plane 3-balls joined at a 2-sphere.  To see this one
makes an embedding.  The embedding of this 3-space can be easily done
in 4-dimensional Euclidean space $\mathbb{R}^{4}$.  In
Fig.~\ref{fig:time_slice_closed_universe} we show an embedding diagram
of a $\theta=\frac{\pi}{2}$ slice of the static Minkowski-Minkowski
closed universe in a 3-dimensional Euclidean space, displaying clearly
the squashed character of the 3-sphere.  One can also make appropriate
identifications between points in the interior and exterior spherical
pieces to turn the space into a projective space.
The causal structure of the resulting spacetime can be shown in a
Carter-Penrose diagram, as in
Fig.~\ref{fig:Carter-Penrose_closed_universe}. We use the hash symbol
\# to represent the connected sum of the spacetime manifold in order
to conserve the conformal structure in the Carter-Penrose diagram of
the total spacetime. We see that it represents a universe in which the
spatial sections are highly squashed 3-spheres, i.e., two copies of
two plane 3-balls joined at a 2-sphere, such that if we include time
the total spacetime is a squashed 3-cylinder, the time line times the
squashed 3-sphere.  A timelike geodesic, or a free-falling particle,
initially moving along the radial coordinate towards increasing values
of it in one half of the spacetime, would reach the domain wall at
$r=R$ at some point,
and then continue until it reaches the center of coordinates at
the other half where it would continue its trajectory into the
antipode point of the wall, and so on.

\newpage

The Minkowski-Minkowski static closed universe is marginally stable.
Indeed, one has from Eq.~(\ref{stabMM}) that $V''(R_{0})=0$.  The
solution is in neutral equilibrium, meaning that for a slight
displacement the thin shell stays at the new radius. This result
confirms the expectations.  as the interior and exterior spacetimes
are both described by a Minkowski, i.e., flat, solution.

This Minkowski-Minkowski static closed universe is a bubble 
universe which is summarized in
Eqs.~(\ref{eq:Mink_metric})-(\ref{eq:pressure_Closed_Universe}) and in
Figs.~\ref{fig:time_slice_closed_universe}
and~\ref{fig:Carter-Penrose_closed_universe}. Moreover, a fundamental
shell is defined as a shell with Minkowski interior with a center and
one of the three basic exterior spacetimes, Minkowski, Schwarzschild,
and Reissner-Nordstr\"om.  Thus the Minkowski-Minkowski static closed
universe we just found completes the search of all the fundamental
shells in the three basic ambient spacetimes, namely,
Minkowski-Minkowski, Minkowski-Schwarzschild, and
Minkowski-Reissner-Nordstr\"om, the latter two having been found
previously.

The Minkowski-Minkowski static closed universe is a representative of
the set of closed universes. It can be compared with other such closed
universes. This will be done later.

\subsubsection{Minkowski-Minkowski static open universe:
A traversable wormhole}

Here we display a Minkowski-Minkowski static open universe as a
solution of general relativity. We rely on the previous results.
We assume that the Minkowski line element is valid for a
region, which we call interior $\mathcal{M}_{\mathrm{i}}$, that goes
from spatial infinity to a radius $R$, i.e., $R\leq
r_\mathrm{i}<\infty$, where $r_\mathrm{i}$ denotes the interior radial
coordinate.  We join this region to another region, which we call
exterior $\mathcal{M}_\mathrm{e}$, where the exterior radial
coordinate is denoted by $r_\mathrm{e}$.  The junction is done at a
common hypersurface $\mathcal{S}$ with circumferential radius
$r_\mathrm{i}=r_\mathrm{e}=R$.  Thus, the whole spacetime is composed
by the two regions plus the common hypersurface, which is a thin
shell.  The common hypersurface $\mathcal{S}$ is assumed to be static.
Assuming the existence of a vector field $n$, normal, at each point,
to the common hypersurface $\mathcal{S}$, we have found that the
solution depends on the orientation of this normal field $n$.  For
each region, $\mathcal{M}_{\mathrm{i}}$ and
$\mathcal{M}_{\mathrm{e}}$, the orientation of the normal is encoded
in a single parameter, namely, one for the interior,
$\xi_{\mathrm{i}}$, and one for the exterior, $\xi_{\mathrm{e}}$. In
both cases, $\xi_{\mathrm{i}}$ and $\xi_{\mathrm{e}}$ can have values
$+1$ or $-1$.  The value $+1$ indicates that the normal points in the
direction of increasing radial coordinate and the value $-1$ indicates
that the normal points in the direction of decreasing radial
coordinate.  The solutions with $\xi_{\mathrm{i}}=\xi_{\mathrm{e}}$
are trivial as the resulting spacetime is simply the full Minkowski
flat universe.  Here, we consider the second non-trivial solution,
i.e., $\xi_{\mathrm{i}}=-1$ and $\xi_{\mathrm{e}}=+1$.  This static
solution represents the case where the normal vector field $n$ points
in the direction of decreasing radial coordinate as seen from the
interior spacetime $\mathcal{M}_{\mathrm{i}}$ and points in the
direction of increasing radial coordinate as seen from the exterior
spacetime $\mathcal{M}_{\mathrm{e}}$. Since $n$ is assumed to point
from $\mathcal{M}_{\mathrm{i}}$ to $\mathcal{M}_{\mathrm{e}}$ this
implies that in the exterior region one also has $R\leq
r_\mathrm{e}<\infty$.  Thus, this solution is composed by two
spatially open Minkowski spacetime regions glued together at the
common boundary $\mathcal{S}$.  This solution then represents a
Riemann flat spacetime everywhere except at $\mathcal{S}$. Overall it
is a static closed Minkowski-Minkowski universe, for which the line
element can be written as, see also
Eqs.~(\ref{eq:Mink_metric_interior})
and~(\ref{eq:Mink_metric_exterior}),
\begin{equation}\label{eq:Mink_metric2}
ds^{2}=-dt^{2}+dr_\mathrm{i}^{2}+r_\mathrm{i}^{2}d\Omega^{2},
\quad R\leq r_\mathrm{i}<\infty\,,
\quad\quad\;\;\;
ds^{2}=-dt^{2}+dr_\mathrm{e}^{2}+r_\mathrm{e}^{2}d\Omega^{2},
\quad R\leq r_\mathrm{e}<\infty\,.
\end{equation}
Now we turn to the properties of the matter shell at $\mathcal{S}$.
In this case, putting $\xi_{\mathrm{i}}=-1$ and $\xi_{\mathrm{e}}=+1$
into Eqs.~(\ref{sig1}) and~(\ref{press1}), the energy density $\sigma$
and the tangential pressure $p$ at the thin shell are given by
\begin{equation}
\sigma=-\frac{1}{2\pi R}\,,
\label{eq:sigma_Wormhole}
\end{equation}
\begin{equation}
p=\;\;\frac{1}{4\pi R}\,.
\label{eq:pressure_Wormhole}
\end{equation}
This solution is then characterized by the presence of a surface layer
or thin shell at radius $R$, separating two Minkowski open halves. The
thin matter shell is composed of a perfect fluid with negative energy
density and is supported by pressure such that it obeys the equation
of state $\sigma+2p=0$. Moreover, from Eqs.~\eqref{eq:sigma_Wormhole}
and \eqref{eq:pressure_Wormhole}, we find that the following
inequalities are verified: $\sigma\leq0$, $\sigma+p\leq0$,
$\sigma+2p\leq0$ and $\sigma\leq\left|p\right|$, therefore, the matter
shell violates the null, weak, strong and dominant pointwise energy
conditions. Since the effective  mass $m$ can be defined by
$\sigma+2p$ and this latter is zero, one has $m=0$. So the shell
yields no total mass $m$ as it should to have Minkowski spacetime
on both sides of the shell.
The volume of this spacetime, in its simplest
personification, i.e., without making identifications
for lare $r$, is infinite.

\begin{figure}[h]
\centering\includegraphics[height=0.17\paperheight]
{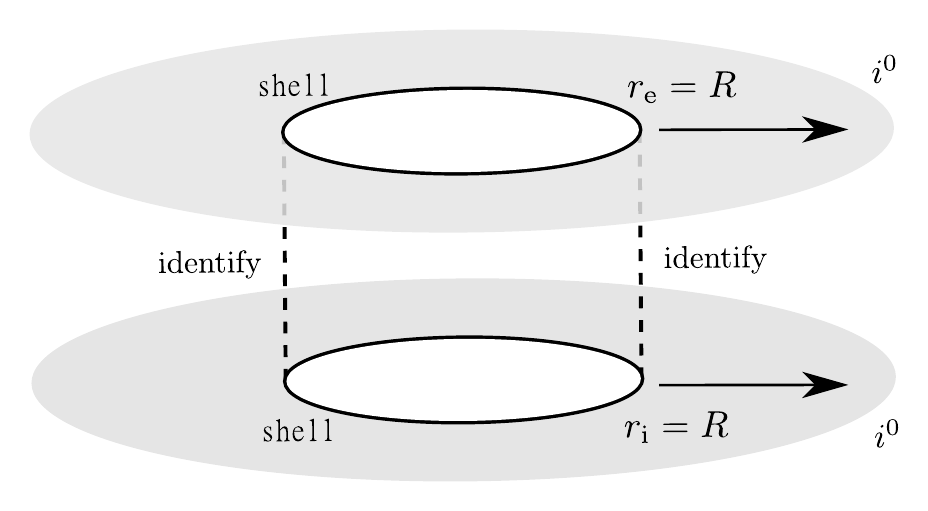}
\caption{\label{fig:time_slice_wormhole}
Embedding diagram of a $t=\text{constant}$ and $\theta=\frac{\pi}{2}$
slice of the Minkowski-Minkowski static open universe, or traversable
wormhole, in 3-dimensional Euclidean space.  The interior coordinate
is $R\leq r_{\mathrm{i}}<\infty$, the exterior coordinate is $R\leq
r_{\mathrm{e}}<\infty$, the radius of the shell is $R$, and the
borders of the circumferences should be identified.
}
\end{figure}

\begin{figure}[h]
\centering\includegraphics[height=0.25\paperheight]
{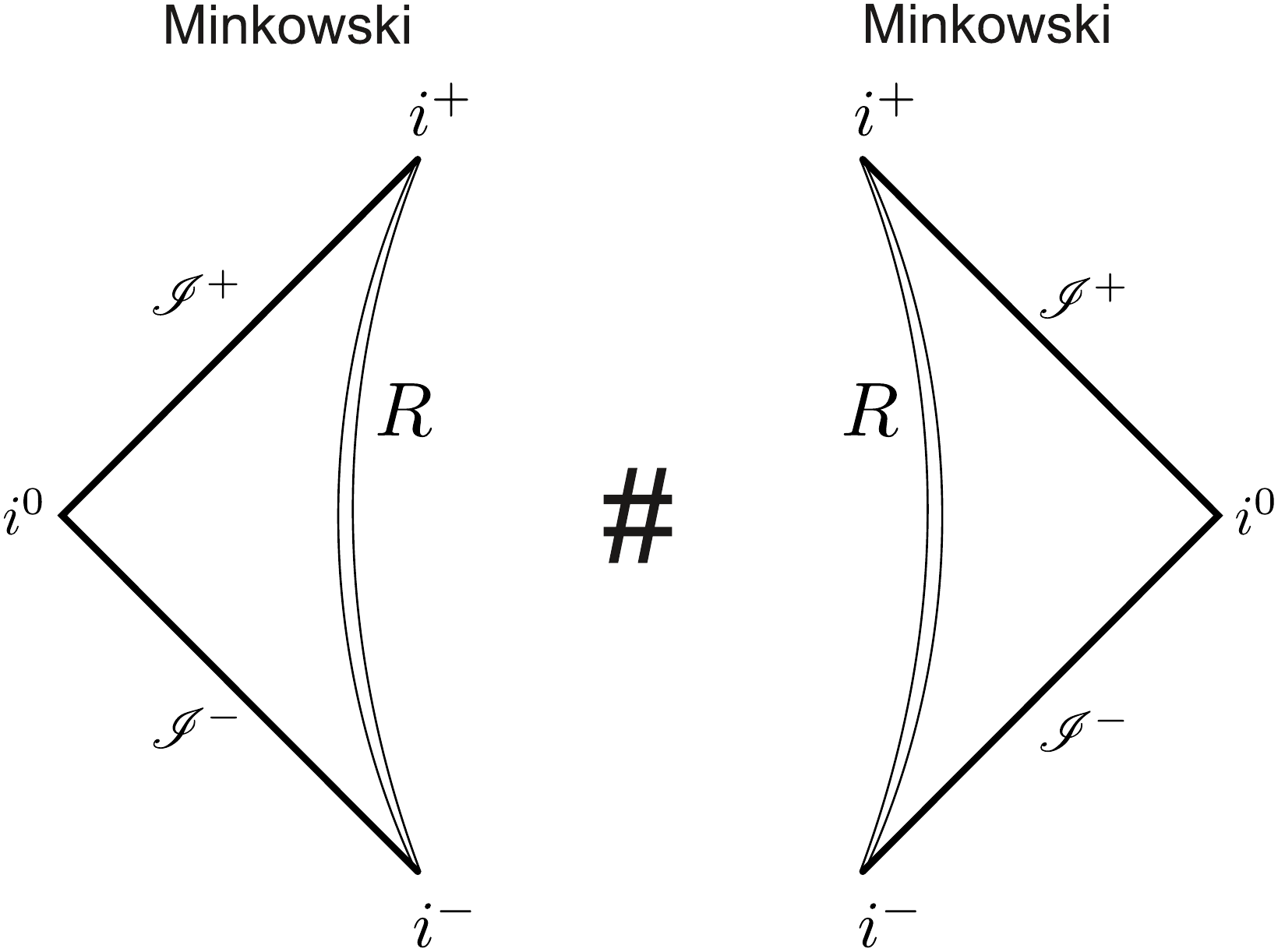}
\caption{
\label{fig:Carter-Penrose_wormhole}
Carter-Penrose diagram of the Minkowski-Minkowski static open
universe, or traversable wormhole.  The hash symbol \# represents the
connected sum of the spacetime manifold. The symbols $i^-$, $i^0$, and
$i^+$ represent past timelike infinity, spatial infinity, and future
timelike infinity, respectively, and the symbols $\mathscr{I}^-$ and
$\mathscr{I}^+$, represent past and future null infinity,
respectively.
}
\end{figure}

The spatial structure and the causal structure are also important to
analyze.  A time slice $t={\rm constant}$ of the spacetime gives that
the 3-space is a universe in which the spatial sections are two copies
of the complements of two plane 3-balls joined at a 2-sphere, the
throat, yielding a non-simply-connected open universe, more precisely,
a traversable wormhole.  To see this one makes an embedding.  The
embedding of this 3-space can be easily done in 4-dimensional
Euclidean space $\mathbb{R}^{4}$.  In
Fig.~\ref{fig:time_slice_wormhole} we show an embedding diagram of a
constant $\theta=\frac{\pi}{2}$ slice of the static
Minkowski-Minkowski open universe, or traversable wormhole.  One can
also make appropriate identifications of the two open sheets, and turn
the space into, e.g., a flat 3-torus, in which case the space is
closed.
The causal structure of the resulting spacetime can be shown in a
Carter-Penrose diagram, as
in Fig.~\ref{fig:Carter-Penrose_wormhole}.  We use the
hash symbol \# to represent the connected sum of the spacetime
manifolds in order to conserve the conformal structure in the
Carter-Penrose diagram of the total spacetime.  We see that it
represents a universe in which the spatial sections are two copies of
the complements of two plane 3-balls joined at a 2-sphere, the throat,
yielding a traversable wormhole, such that if we include time, the
total spacetime has the topology $\mathbb{R}\times\Sigma$, where
$\Sigma$ is a 3-manifold with nontrivial topology, whose boundary
$\partial\Sigma\sim S^{2}$.  A causal geodesic, or a free-falling
particle, initially moving in the direction of decreasing radial
coordinate in one half of the spacetime, would reach the shell at
$r=R$ and then continue until it reaches infinity at the other sheet
of the wormhole.

\newpage

The Minkowski-Minkowski static open universe, i.e.,
the Minkowski-Minkowski
traversable wormhole, is marginally stable.
Indeed, one has from Eq.~(\ref{stabMM}) that $V''(R_{0})=0$.  The
solution is in neutral equilibrium, meaning that for a slight
displacement the thin shell stays at the new radius. This result
confirms the expectations.  as the interior and exterior spacetimes
are both described by a Minkowski, i.e., flat, solution.

This Minkowski-Minkowski static open universe is a traversable
wormhole which is summarized in
Eqs.~(\ref{eq:Mink_metric2})-(\ref{eq:pressure_Wormhole}) and in
Figs.~\ref{fig:time_slice_wormhole}
and~\ref{fig:Carter-Penrose_wormhole}.  This open universe is an
exotic shell spacetime rather than a fundamental shell spacetime, as
its interior does not contain a center or origin, instead the interior
opens up to infinity.  This further possibility for a shell spacetime,
i.e., that its interior opens up to infinity, implies that the
spacetime is a traversable wormhole spacetime, and thus the matter
properties of the shell must be exotic since they necessarily violate
the energy conditions.  The study of the Minkowski-Minkowski open
spacetime, or traversable wormhole, introduces the prospect of
analyzing all possible exotic shells, i.e., shells for which the
Minkowski interior has no center, in the other two basic ambient
spacetimes, namely, Schwarzschild and Reissner-Nordstr\"om spacetimes.

The Minkowski-Minkowski static open universe is a representative of
the set of traversable wormholes. It can be compared with other
such open universes and traversable wormholes.
This will be done later.

\subsection{Minkowski-Minkowski universes: One concept with two sides}

The two, close and open, Minkowski-Minkowski spacetimes demonstrate
that they can be seen as complementary to each other, i.e., they are
two sides of the same concept.  The concept, i.e., a collection of two
Minkowski spacetimes together, that when cut into spherical
regions yield on one side  a
closed universe, a bubble universe, 
and on the other side  an open universe
which is a traversable wormhole.
The formalism presented in analyzing the two Minkowski-Minkowski
universes is well suited to show this point.
Indeed, from an algebraic point of view, one side is given by
$\frac12(\xi_{\rm i}-\xi_{\rm e})=1$, the other side is given by
$\frac12(\xi_{\rm i}-\xi_{\rm e})=-1$, where $\xi_{\rm i}$ and
$\xi_{\rm e}$ are the characteristics of the interior and exterior
normals to the shell, respectively.
This algebraic side appears clearly in the evaluation of
the matter properties as displayed in
Eqs.~(\ref{sig1}) and~(\ref{press1}).
More formally, to implement the idea of a Minkowski-Minkowski closed
universe, i.e., a bubble universe,
and a Minkowski-Minkowski open universe, i.e., a traversable
wormhole, one uses the equations of general relativity together with
the appropriate thin shell formalism. For a shell with a Minkowski
interior with a center, when the normal to the shell in the exterior
region points towards decreasing $r$, i.e., $\frac12(\xi_{\rm
i}-\xi_{\rm e})=1$, one finds the Minkowski-Minkowski closed universe,
made of two 3-dimensional flat balls, or sheets, that are joined
at some domain wall, i.e., a 2-sphere shell with matter, to make a
Minkowski-Minkowski bubble universe.  For a shell with a Minkowski
open interior, when the normal to the shell points towards increasing
$r$ in the Minkowski exterior, i.e., $\frac12(\xi_{\rm i}-\xi_{\rm
e})=-1$, one finds the Minkowski-Minkowski open universe, made of two
3-dimensional flat open infinite sheets that are joined at some
throat, to make a Minkowski-Minkowski traversable
wormhole.
From a matter point of view the two universes show a form of
complementarity, as for $\frac12(\xi_{\rm i}-\xi_{\rm e})=1$ the matter
obeys the energy conditions while for $\frac12(\xi_{\rm i}-\xi_{\rm
e})=- 1$ the matter violates the energy conditions.
From a geometrical point of view, the two sides of the concept appear
when one picks up a Minkowski spacetime and at constant time cuts a
ball in it, to obtain two spaces, namely, a 3-dimensional ball with a
flat inside, and an infinite extended 3-dimensional flat space with a
hole, which is the complement of the ball. Then one picks up another
Minkowski spacetime and do the same, to get a second ball and a second
infinite extended flat space with a hole.  One side  is
given if one joins the two 3-dimensional balls along a 2-sphere, a
shell containing matter, to obtain a single 3-space that including
time makes altogether a static closed Minkowski-Minkowski universe, a
bubble universe. The other side is given if one joins the
two complements, i.e., the two infinite extended 3-dimensional
flat spaces with a hole in each, along a 2-sphere, a shell containing
matter, to obtain a different single 3-space that including time
makes altogether another Minkowski-Minkowski universe, which is a
traversable wormhole. Comparison of 
Fig.~\ref{fig:time_slice_closed_universe}
with
Fig.~\ref{fig:time_slice_wormhole} for a spatial geometrical
representation of the bubble universe and the traversable wormhole,
respectively, displays the complementarity of the two spaces clearly,
which can be further strengthened with the comparison of the spacetime
drawings in the form of Carter-Penrose diagrams, given in
Fig.~\ref{fig:Carter-Penrose_closed_universe} and in
Fig.~\ref{fig:Carter-Penrose_wormhole}, respectively.
From a stability point of view it is also interesting that
both spacetimes
are marginally stablee, showing thus here some form
of neutral complementarity.

So, that the two Minkowski-Minkowski spacetimes demonstrate that they
can be seen as complementary to each other, i.e., they are two sides
of the same concept, is clear. It can be raised the point
that the bubble universe
has matter that obeys the energy conditions, whereas the traversable
wormhole has matter that does not obey those conditions.
This is obviously true, but there is no
real problem with it.
In an early era of the Universe, when quantum gravity
dominates, there is really no obeyance  to the energy conditions and
the closed and open universes, created as bubble universes with domain
walls and traversable wormholes with throats out of the spacetime foam
must coexist together. Some kind of inflation would grow these objects
to macroscopic dimensions turning them into new structures inhabiting
the Universe itself, showing that 
bubble universes and traversable wormholes are distinct but
connected objects, some obeying the energy
conditions and others
not.



\section{Einstein static closed universe and Friedmann static
hyperbolic open universe}
\label{sec:closedandopenEFuniverses}

\subsection{Einstein and
Friedmann static universes: Formal solutions and stability}

\subsubsection{Solutions}

Two paradigmatic solutions of the theory of general relativity for
static universes are the Einstein and the hyperbolic Friedmann
spacetimes.  These two solutions have various resemblances with the
open and closed Minkowski-Minkowski universes studied in the previous
section, and we will present them in a form suited for comparing their
properties.  Consider then the Einstein field equations with a
non-vanishing cosmological constant $\Lambda$,
\begin{equation}
G_{\alpha\beta}+\Lambda g_{\alpha\beta}=8\pi
T_{\alpha\beta}\,,\label{eq:EFE2forEeF}
\end{equation}
where $G_{\alpha\beta}=\mathcal R_{\alpha\beta}-
\frac{1}{2}g_{\alpha\beta}\mathcal{R}$
is the Einstein tensor, $\mathcal R_{\alpha\beta}$
and $\mathcal{R}$ are
the Ricci tensor and Ricci scalar, respectively, $g_{\alpha\beta}$
is the spacetime metric, and $T_{\alpha\beta}$ is the stress-energy
tensor.

One assumes a static, homogeneous and isotropic spacetime
and, in addition, one supposes that $T_{\alpha\beta}$ corresponds
to a perfect fluid which has energy density $\rho$ and
vanishing pressure
$p$, i.e., a dust-like fluid.
With these assumptions,
the solution of the field equations~\eqref{eq:EFE2forEeF}
in spacetime spherical coordinates $\left(t,r,\theta,\varphi\right)$
is given by the line element
\begin{equation}
ds^{2}=-dt^{2}+dr^{2}+R^{2}
\left[\frac{1}{\sqrt k}
\sin
\left(\sqrt{k}\frac{r}{R}\right)\right]^2d\Omega^{2}\,,
\label{eq:line_element_EFgeneral}
\end{equation}
where $t$ is the time coordinate, $r$ is the radial coordinate,
$d\Omega^{2}\equiv d\theta^{2}+\sin^{2}\theta d\varphi^{2}$, with
$\theta$ and $\varphi$ being the spherical angular coordinates,
$0\leq\theta\leq\pi$ and $0\leq\varphi\leq2\pi$, $R$ is a positive
constant scale factor representing a characteristic radius of the
universe, and $k$ is related with the Ricci curvature 
scalar by $\mathcal{R}=\frac{6k}{R^{2}}$ and may take the values
$k=1,0,-1$.
Furthermore, the field equations~\eqref{eq:EFE2forEeF} also 
give expressions
for the energy density of the fluid and for the cosmological constant, namely,
\begin{align}
\rho & =\frac{k}{4\pi R^{2}}\,,
\label{eq:stability_sigma_value_generalEF}\\
\Lambda & =\frac{k}{R^{2}}\,,
\label{eq:stability_pressure_value_generalEF}
\end{align}
and $p=0$. From Eqs.~\eqref{eq:stability_sigma_value_generalEF}
and \eqref{eq:stability_pressure_value_generalEF}
one finds that
\begin{equation}
\rho - \frac{\Lambda}{4\pi}=0\,,
\label{sum}
\end{equation}
and so $\Lambda$ counteracts the gravitational pulling effects of $\rho$.
Noticing that the term $\Lambda g_{\alpha\beta}$ can be thought as
a perfect fluid contribution to the stress-energy tensor with energy
density $\bar{\rho}=\frac{\Lambda}{8\pi}$ and pressure
$\bar{p}=-\frac{\Lambda}{8\pi}$, this static homogeneous universe
solution can then be seen as a solution of a two fluid system,
one fluid with energy density $\rho=\frac{\Lambda}{4\pi}$ and pressure
$p=0$, and the other fluid, a vacuum fluid, with energy density
$\bar{\rho}=\frac{\Lambda}{8\pi}$ and pressure
$\bar{p}=-\frac{\Lambda}{8\pi}$,
such that $\rho+\bar{\rho}+3\bar{p}=0$.
Besides the trivial case, i.e., the Minkowski universe  which
has $k=0$, there are
two possible universes at this juncture, the Einstein static closed
universe which has $k=1$, and the Friedmann static open universe which
has $k=-1$.

\subsubsection{Linearized stability analysis for Einstein and
Friedmann static universes}

An important question regarding the static
homogeneous universes, i.e.,
the Einstein and  hyperbolic
Friedmann static
universes, is if these are stable under perturbations. Here
we will discuss the linear stability of these cosmological solutions
by analyzing the equation of motion of the universe near the static
configuration. The result for the Einstein universe is well known,
whereas the stability of the static
hyperbolic Friedmann universe
is less known.

To study the linear stability of the static Einstein and Friedmann
universes we have to find the evolution equation for the scalar
factor $R$ and analyze the behavior of these solutions as we perturb
the spacetime.
The analysis can be done in a unified way by making use of
the parameter $k$.
From the general relativity field equations we find
the Friedmann equation, namely,
\begin{equation}
\dot{R}^{2}+ V(R) =0\,,
\label{adots}
\end{equation}
where we have introduced the potential
\begin{equation}
V(R)=k-\frac{8\pi\rho+\Lambda}{3}R^{2} \,,
\label{eq:stability_potentiala}
\end{equation}
and the scale factor $R$ is now
a function of the time coordinate $t$, $R=R(t)$,
and a dot represents derivative
with respect to it,
and again
$k$ represents the sectional curvature of constant
time slices such that, $k=+1$ for the closed universe,
$k=0$ for the flat universe, and $k=-1$
for the open universe.
Note anew that a universe is stable
if, and only if, the potential $V(R)$ is at a local
minimum, i.e., if $V'(R)=0$ and
$V''(R)\geq0$, with the equality providing the
marginal neutral case, where a
prime denotes the derivative with respect to $R$.
Thus, we have to calculate the matter properties
and its derivatives.
All these properties are functions of the shell radius $R$,
namely, $\rho=\rho(R)$,
$p=p(R)$,
$\rho'=\rho'(R)$, $p'=p'(R)$, and the
radius itself is a function of time $R(t)$.
The
conservation equation,
which can be taken from the field equations,
is $\dot\rho+\frac{3\dot R}{R}\rho=0$, so that
the equation for $\rho'$, 
where prime denotes the derivative with respect to $R$, is
\begin{equation}
\rho'=-3\frac{\rho}{R}\,.
\label{eq:stability_rho_prime}
\end{equation}
We have now
to analyze the derivatives of
the potential $V(R)$ given in
Eq.~\eqref{eq:stability_potentiala} at the static configuration.
From
Eq.~\eqref{eq:stability_potentiala} we get
$V'(R)=
-\frac13 \left(8\pi\rho'\right)R^{2}
-\frac23\left( 8\pi\rho+\Lambda \right)\,R$,
where we assume that $\Lambda$ is a constant.
Taking the derivative of it we get
$V''(R)=
-\frac13 \left(8\pi\rho''\right)R^{2}
-\frac43 \left(8\pi\rho'\right)R
-\frac23\left(8\pi\rho+\Lambda
\right)$.
Now, if we introduce
Eq.~(\ref{eq:stability_rho_prime})
into 
$V'(R)$ we obtain
$V'(R)=\frac23\left(4\pi\rho-\Lambda\right)R$.
Simplifying also $V''(R)$ we obtain
$V''(R)=-\frac23\left(8\pi\rho+\Lambda\right)$.
In brief, the derivatives of the potential
$V(R)$ are
\begin{equation}
V'(R)=
\frac23\left(4\pi\rho-\Lambda\right)R\,,
\label{eq:stability_V_primeEF}
\end{equation}
and
\begin{equation}
V''(R)=
-\frac23\left(8\pi\rho+\Lambda
\right)\,,
\label{Vdashdasha}
\end{equation}
where we assume that $p=0$ throughout, i.e., the
cold generic equation $p=p(\rho)$ is the trivial one,
so
here
$\eta\left(\rho\right)\equiv\frac{\partial p}{\partial\rho}$
is zero,
$\eta\left(\rho\right)=0$.

Following the usual reasoning, we linearize
Eq.~\eqref{adots} around a static solution. Defining
$R_{0}$ as the radius of the static universe and
assuming the potential $V$ to be a differentiable function at $R_{0}$,
we can expand the potential~\eqref{eq:stability_potentiala} around
$R_{0}$ as
\begin{equation}
V(R)=V(R_{0})+V'(R_{0})
\left(R-R_{0}\right)+\frac{1}{2}V''(R_{0})
\left(R-R_{0}\right)^{2}\,,
\label{eq:stability_potential_expansionEF}
\end{equation}
plus higher order terms of
$\mathcal{O}\left[\left(R-R_{0}\right)^{3}\right]$.
Now, a universe with radius $R_{0}$ is stable
if, and only if, the potential, $V(R)$, is at a local
minimum, i.e., if $V'(R_{0})=0$ and
$V''(R_{0})\geq0$, with the equality providing the
marginal neutral case.  
The static
solutions found in the previous sections are characterized by the
following expressions for the energy density and
the cosmological constant of the fluid,
$\rho  =\frac{k}{4\pi R_0^2}$
and 
$\Lambda=\frac{k}{R_0^2}$, see
Eqs.~(\ref{eq:stability_sigma_value_generalEF})
and~(\ref{eq:stability_pressure_value_generalEF}),
with the fluid pressure $p$ being zero, $p=0$.
Then putting
Eq.~\eqref{eq:stability_sigma_value_generalEF} into
Eq.~\eqref{eq:stability_potentiala} we find $V(R_{0})=0$, as
expected.
Substituting it into Eq.~\eqref{eq:stability_V_primeEF}
and evaluating at the static solution we find
$V'(R_{0})=0$. 
Evaluating $V''(R)$ at the static solution, $R=R_{0}$,
we find
$V''(R_{0})=-\frac{2k}{R_{0}^{2}}$. 
In brief,
defining $R_{0}$
as the value of the scale factor of the static Einstein and hyperbolic
Friedmann universes, expanding the potential $V(R)$ around
$R_{0}$, Eq.~\eqref{eq:stability_potential_expansionEF}, we find that
\begin{equation}
V'(R_{0})=0\,,
\end{equation}
and
\begin{equation}
V''(R_{0})=-\frac{2k}{R_{0}^{2}}\,.
\label{stabEF}
\end{equation}

Gathering these calculations, we conclude that, besides the trivial
case, i.e., the Minkowski universe which has $k = 0$ and is trivially
neutraly stable, there is the Einstein static closed universe which
has $k = 1$ and so is unstable, and the Friedmann static open universe
which has $k = -1$ and so is stable.
The 
instability
of the $k = 1$ Einstein static closed universe
means that if we slightly displace
the scale radius $R$ towards larger or smaller values,
the universe will expand in the former displacement
or collapse in the latter displacement,
and the 
stability
of the $k = - 1$ Friedmann static closed universe
means that if we slightly displace
the scale radius $R$ towards larger or smaller values,
the universe will get back to the initial value
$R$.

\subsection{
Einstein and Friedmann static universes: Geometry
and physics}

\subsubsection{Einstein static closed universe}

The Einstein universe is a solution of the general theory of
relativity for a dust source with energy density $\rho$, pressure $p$
equal to zero, a positive cosmological constant $\Lambda$, and
positive curvature, $k=1$.  In spacetime spherical coordinates
$(t,r,\theta,\varphi)$ it is characterized by the line element given
in Eq.~(\ref{eq:line_element_EFgeneral}) with $k=1$, i.e.,
\begin{equation}
ds^{2}=-dt^{2}+
d{r}^{2}+R^2\sin^{2}\left(\frac{r}{R}\right)\,
d\Omega^{2}
\,,
\label{eq:line_element_EinsteinU}
\end{equation}
where $t$ is the time coordinate, $r$ is the radial coordinate with
$0\leq r\leq\pi\,R$, and $d\Omega^{2}\equiv d\theta^{2}+\sin^{2}\theta
d\varphi^{2}$, with $\theta$ and $\varphi$ being the spherical angular
coordinates, $0\leq\theta\leq\pi$ and $0\leq\varphi\leq2\pi$.  In
addition, $R$ is a positive scale factor which here is a  constant,
and which
gives the characteristic radius of the
universe.  The Ricci scalar for the Einstein universe is given
by $\mathcal{R}=\frac6{R^2}$. This solution then represents
a static spacetime, a 3-dimensional sphere,
and so is
a closed universe.
Now we turn to the properties of the matter in the Einstein universe.
Assuming a perfect fluid made of dust, i.e., the matter has
energy density $\rho$ and pressure $p=0$,
the Einstein field equations with cosmological constant $\Lambda$
for the line element
given in Eq.~(\ref{eq:line_element_EinsteinU})
yield
\begin{equation}
\rho  =\frac{1}{4\pi R^2}\,,
\end{equation}
\begin{equation}
\Lambda  =\frac{1}{R^{2}}\,,
\end{equation}
see Eqs.~(\ref{eq:stability_sigma_value_generalEF}) and
(\ref{eq:stability_pressure_value_generalEF}) with $k=1$.
Note that $\rho - \frac{\Lambda}{4\pi}=0$, see Eq.~(\ref{sum}),
and so $\Lambda$ being positive is repulsive everywhere
and thus 
assumes the function of a pressure
that acts against the gravitational pull of the matter
specified by $\rho$.
This system can be seen as a two fluid
system, one fluid with energy density $\rho$, the other
fluid, a vacuum fluid, with 
energy density ${\bar \rho}=\frac{\Lambda}{8\pi}$ and pressure
${\bar p}=-\frac{\Lambda}{8\pi}$, such that
$\rho+\bar{\rho}+3\bar{p}=0$.
All the matter energy
conditions are satisfied.
The volume of this universe is $V=2\pi^2R^3$ and its mass is
is $m=2\pi^2R^3\rho$.

\begin{figure}[h]
\centering\includegraphics[height=0.21\paperheight]
{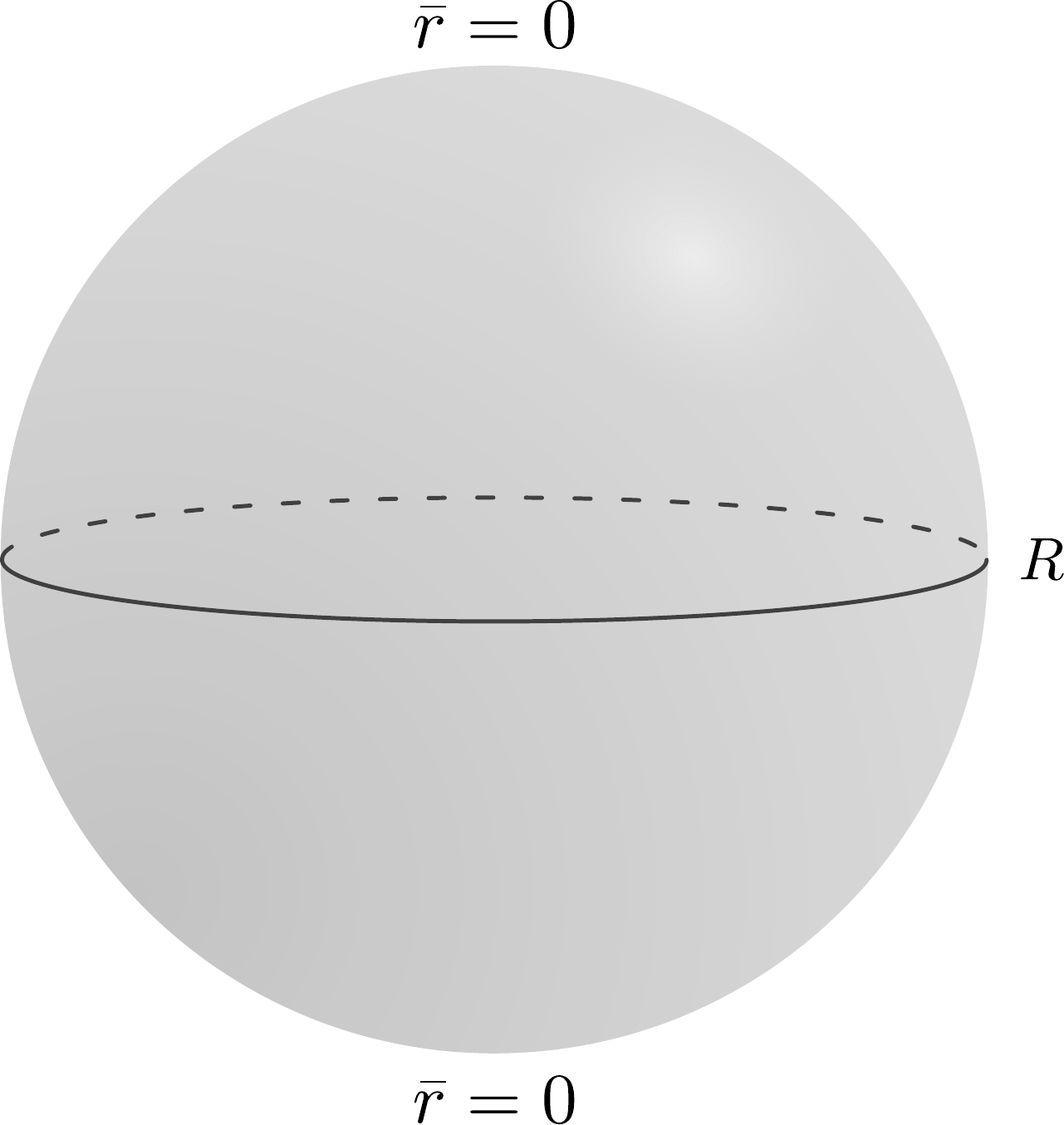}
\caption{\label{fig:time_slice_Einstein_universe}
Embedding diagram of a $t=\text{constant}$ and $\theta=\frac{\pi}{2}$
slice of the Einstein static closed universe in 3-dimensional
Euclidean space. The radial coordinate $\bar r$
related to the area defined by it, namely, ${\bar
r}=R\sin\left(\frac{r}{R}\right)$, is the radial coordinate
used in the diagram.  This coordinate
runs from 0 at one pole, to $R$ at the equatior, and then back to 0 at
the other pole, with $R$ being the characteristic radius of the
Einstein universe.
}
\end{figure}

\begin{figure}[h]
\centering\includegraphics[height=0.25\paperheight]
{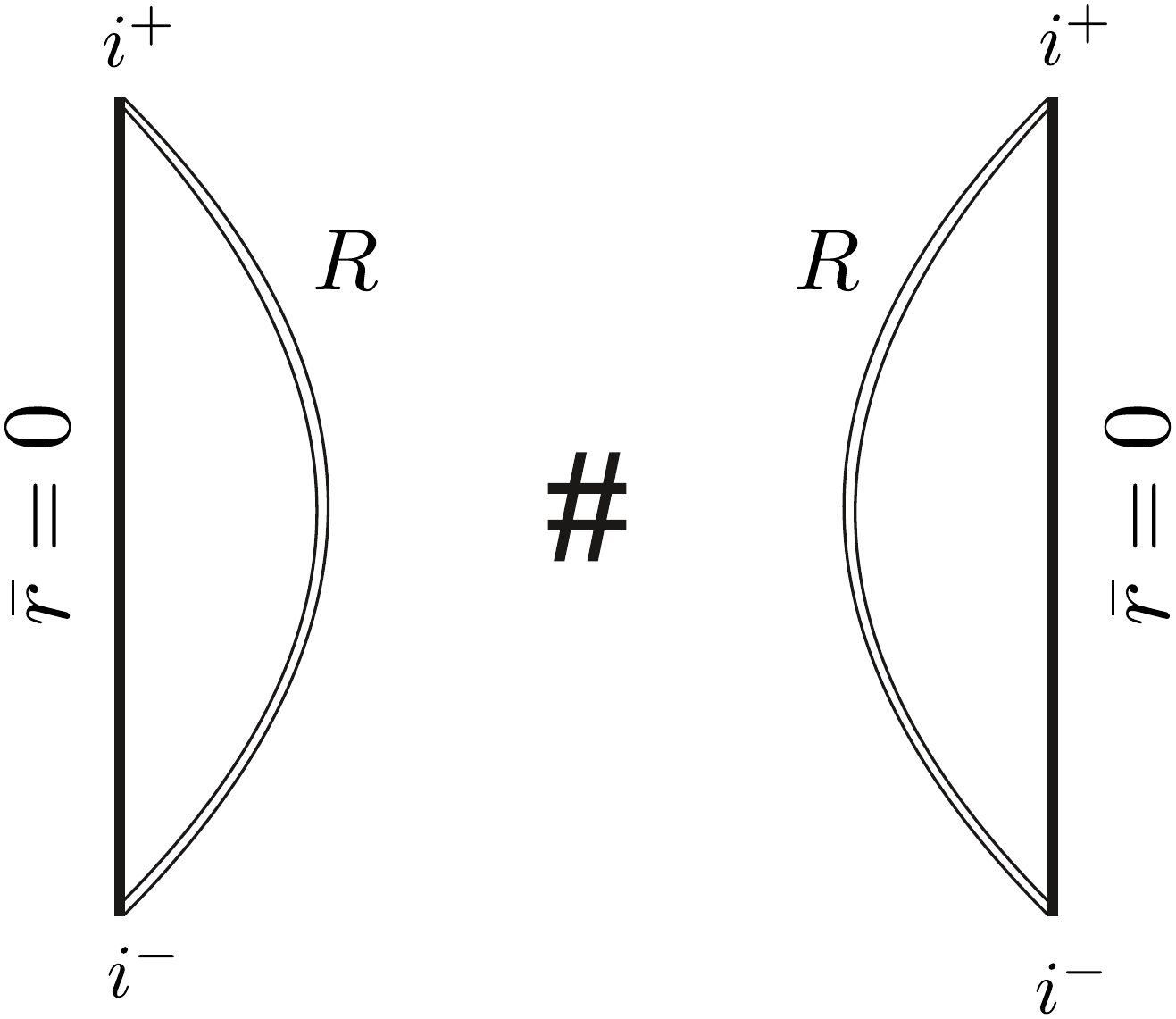}
\caption{\label{fig:Carter-Penrose_Einstein_universe}
Carter-Penrose diagram of the Einstein static closed universe. The
vertical lines represent the two poles of the sphere.  The radial
coordinate $\bar r$ related to the area defined by it, namely, ${\bar
r}=R\sin\left(\frac{r}{R}\right)$, is the radial coordinate used in
the diagram.  One pole is situated at the origin with ${\bar
r}=0$. The other pole has also ${\bar r}=0$. The two lines denoted by
$R$ constitute the equatior.  The hash symbol \# represents the
connected sum of the spacetime manifold. The symbols $i^-$ and $i^+$
represent past and future causal infinity, respectively.
}
\end{figure}

The spatial and causal structure of
the spacetime
can also be presented.
Considering a slice of constant time of the spacetime,
$t=\rm{constant}$, we find that the 3-space is diffeomorphic to a
3-sphere. To show this, we can embed the 3-space in 4-dimensional
Euclidean space, $\mathbb{R}^{4}$. Defining the Euclidean spatial
coordinates $(w,x,y,z)$ as $w=R\cos\frac{r}{R}$,
$x=R\sin\frac{r}{R}\sin\theta\cos\phi$,
$y=R\sin\frac{r}{R}\sin\theta\sin\phi$, and
$z=R\sin\frac{r}{R}\cos\theta$, the line element of the embedded
surface is given by $ds^2=dw^2+dx^2+dy^2+dz^2$, and the surface
verifies the equation $w^2+x^2+y^2+z^2=R^2$, showing that indeed it
can be regarded as a 3-sphere in $\mathbb{R}^4$.
To visualize the embedding one makes a $\theta=\frac{\pi}{2}$ slice,
i.e., $z=0$ in the Euclidean coordinates.  In
Fig.~\ref{fig:time_slice_Einstein_universe} we show such an embedding
for the static
spherical Einstein universe.
By making appropriate identifications
between points in the two hemispheres, the spherical space turns
into a projective spherical space also called an elliptical space.
In
Fig.~\ref{fig:Carter-Penrose_Einstein_universe} we show
the causal structure of the resulting spacetime in a Carter-Penrose
diagram.  The Einstein
universe, a static spacetime, models a universe with spherical spatial
sections such that if we include time the total spacetime is a
3-cylinder, $\mathbb{R}\times S^3$.
A timelike geodesic, or a
free-falling particle, initially moving from $r=0$
in the direction of increasing
radial coordinate  would reach the other pole
at $r=\pi R$ and then continue until it reaches 
back the center of coordinates and so forth.

The Einstein static closed universe is unstable.  Indeed, from
Eq.~(\ref{stabEF}) one has that for $k=1$, $V''(R_{0})<0$, recovering
the well known result that the Einstein static closed universe is
unstable under perturbations.  This result confirms the expectations.
For the static Einstein universe a small increase in the radius of the
universe means less gravitational field due to matter and more
cosmological repulsion field from $\Lambda$, so it is a runaway
expanding unstable solution, and reversing the argument for a small
decrease in the radius one finds a runaway contracting unstable
solution. So, although it obeys the energy conditions and is a priori
not problematic, it is unstable, giving rise
to an expanding bubble universe.

This Einstein static closed universe is well known.  It was extremely
important in initiating the science of cosmology.  The requirement
that the boundary conditions on the gravitational field should be
finite and consistent led to a closed universe, which in turn was also
relevant to make the point that general relativity could be Machian,
i.e., that geometry and inertia would arise solely from
matter.
The requirement that the universe should be static, as was thought at
the time, yielded a new constant to physics, the cosmological
constant. The corresponding cosmological term added to
the original general theory of relativity provided
in turn 
the first modified gravitational theory.

We can now make a comparison between the Minkowski-Minkowski static
closed universe and the Einstein universe. Although the two universes
are, of course, totally distinct solutions of the general
relativistic field equations, there are differences
and also striking similarities between them.
In relation to the matter properties, the
Minkowski-Minkowski closed universe is highly nonuniform, it is vacuum
everywhere except at a thin shell with circumferential radius $R$,
made of a perfect fluid with a positive energy density $\sigma$ and a
positive, repulsive, pressure $p$ to hold it static against
gravitational collapse or expansion. The Einstein universe, with
characteristic radius $R$, is uniform, permeated by a fluid with a
positive energy density $\rho$ and a repulsive cosmological constant
to hold it static against gravitational collapse or expansion.  Thus,
both universes obey the energy conditions,
they have positive densities and have some
form of pressure, negative tangential shell pressure in one case and
positive cosmological constant pressure in the other case, to hold
them static.  In relation to the geometric and causal properties, one
can compare the figures drawn, namely, a $t=\text{constant}$ and
$\theta=\frac{\pi}{2}$ slice of the Minkowski-Minkowski closed
universe and the corresponding Carter-Penrose diagram shown in
Figs.~\ref{fig:time_slice_closed_universe} and
~\ref{fig:Carter-Penrose_closed_universe}, respectively, and a
$t=\text{constant}$ and $\theta=\frac{\pi}{2}$ slice of the Einstein
closed universe and the corresponding Carter-Penrose diagram shown in
Figs.~\ref{fig:time_slice_Einstein_universe}
and~\ref{fig:Carter-Penrose_Einstein_universe}.
The comparison leads to the conclusion that the two universes have an
evident similar 
geometrical structure.  The Minkowski-Minkowski closed universe
models a universe with squashed spherical spatial sections such that
the total spacetime is a squashed 4-cylinder with 
$\mathbb{R}\times S^{3}$, with the possibility of
further identifications 
in $S^{3}$.  The Einstein closed universe models a
universe with spherical spatial sections such that the total spacetime
is a 4-cylinder $\mathbb{R}\times S^{3}$, with the possibility of
further identifications 
in $S^{3}$.  Both universes have thus
spherical spatial topology and similar causal structures.
The Minkowski-Minkowski closed universe is the result of compressing,
in a sense, the evenly distributed matter of the Einstein universe
into a thin shell leaving the rest of the spacetime empty.  The
Minkowski-Minkowski closed universe is stable, marginally, and the
Einstein closed universe is unstable, so, since there are no
topological obstructions, a possible endpoint of the Einstein closed
universe, if perturbed at constant universe radius,
could be the Minkowski-Minkowski closed universe.

\subsubsection{Friedmann static hyperbolic open universe: A
failed wormhole}

The Friedmann static universe is a solution of the general theory
of relativity for a dust source with negative energy density $\rho$,
pressure $p$ equal to zero, a negative cosmological constant
$\Lambda$, and negative curvature $k = -1$.
In spacetime
hyperspherical coordinates $(t,r,\theta, \varphi)$
it is characterized
by the line element
given
in Eq.~(\ref{eq:line_element_EFgeneral}) with $k=-1$, i.e.,
\begin{equation}
ds^{2}=-dt^{2}+
d{r}^{2}+R^2\sinh^{2}\left(\frac{r}{R}\right)\,
d\Omega^{2}
\,,
\label{friedmanstaticmetric}
\end{equation}
where $t$ is the time coordinate, $r$ is the radial coordinate with
$0\leq r<\infty$, and $d\Omega^{2}\equiv d\theta^{2}+\sin^{2}\theta
d\varphi^{2}$, with $\theta$ and $\varphi$ being the spherical angular
coordinates, $0\leq\theta\leq\pi$ and $0\leq\varphi\leq2\pi$.  In
addition, $R$ is a positive
scale factor which here is a constant,
and which gives the characteristic radius of the universe.  The Ricci
scalar for the Friedmann universe is given by
$\mathcal{R}=-\frac6{R^2}$, so the Friedmann universe is a negative
constant curvature spacetime. Clearly it is a static hyperbolic
spacetime, and so an open universe.
Now we turn to the properties of the matter in the
static Friedmann universe.
Assuming a perfect fluid made of dust, i.e., the matter has
energy density $\rho$ and pressure $p=0$,
the Einstein field equations with negative cosmological constant $\Lambda$
for the line element
given in Eq.~(\ref{friedmanstaticmetric})
yield
\begin{equation}
\rho  =-\frac{1}{4\pi R^2}\,,
\end{equation}
\begin{equation}
\Lambda  =-\frac{1}{R^{2}}\,,
\end{equation}
see Eqs.~(\ref{eq:stability_sigma_value_generalEF}) and
(\ref{eq:stability_pressure_value_generalEF}) with $k=-1$.
Note that
$\rho - \frac{\Lambda}{4\pi}=0$, see Eq.~(\ref{sum}),
and so $\Lambda$
being negative is attractive everywhere,
and thus assumes the function of a tension
that acts against the gravitational push of the matter specified by
a negative energy density $\rho$.
This system can be seen as a two fluid
system, one fluid with negative energy density$\rho$, the other
fluid, a vacuum fluid, with 
negative energy density ${\bar \rho}=\frac{\Lambda}{8\pi}$ and tension
${\bar p}=-\frac{\Lambda}{8\pi}$
such that $\rho+\bar{\rho}+3\bar{p}=0$.
The matter energy
conditions are violated.
The volume of this hyperbolic
universe, in its open form, is $V=\infty$ and 
its mass is also infinite.

\begin{figure}[h]
\centering\includegraphics[height=0.2\paperheight]
{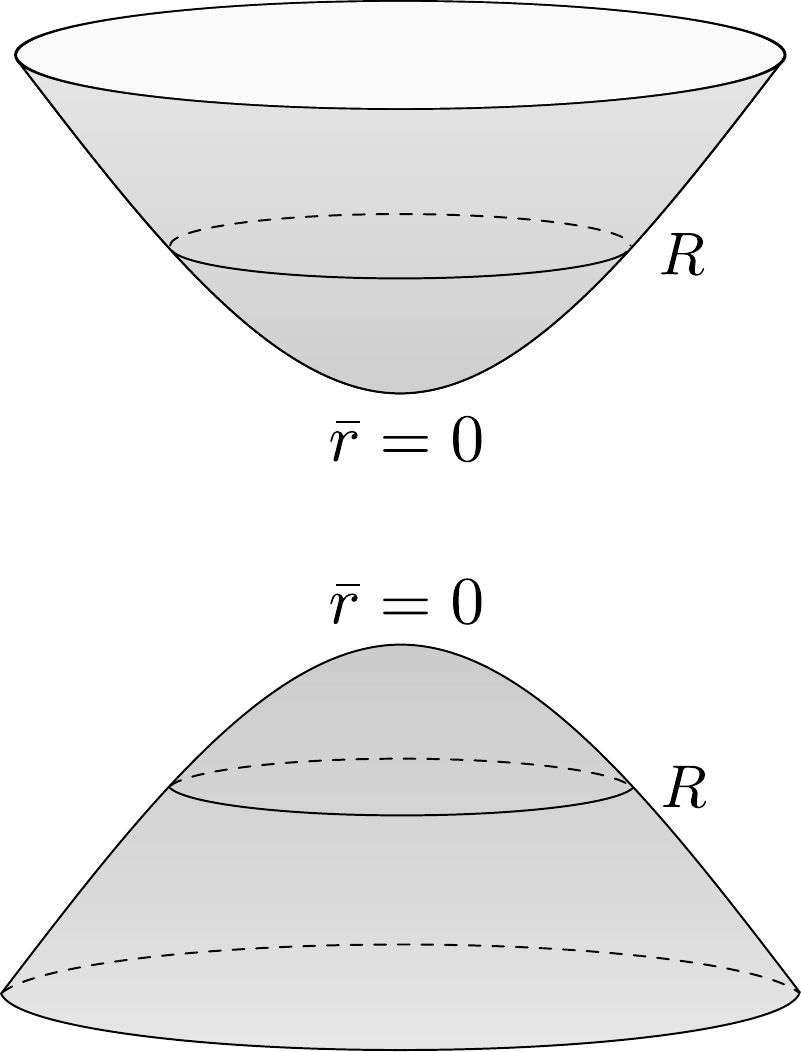}
\caption{\label{fig:time_slice_Friedmann_wormhole}
Embedding diagram of a $t=\text{constant}$ and $\theta=\frac{\pi}{2}$
slice of the Friedmann static open universe in 3-dimensional
Minkowski space. The radial coordinate $\bar r$
related to the area defined by it, namely, ${\bar
r}=R\sinh\left(\frac{r}{R}\right)$, is the radial coordinate
used in the diagram.  This coordinate
runs from 0 at one pole, to infinity, passing through $R$ a some
point, and the same at
the other pole, it runs from 0, to infinity, passing through $R$ a some
point, with $R$ being the characteristic radius of the
Friedmann static universe.
The poles are at
the origin which is common to both branches,
sheets, or branes of the spacetime, although this cannot be seen
explicit in the embedding. The branches are separated,
and the solution is a failed wormhole. Indeed, there are two
disconnected spacetimes almost connected at ${\bar r}=0$.
}
\end{figure}

\begin{figure}[h]
\centering\includegraphics[height=0.25\paperheight]
{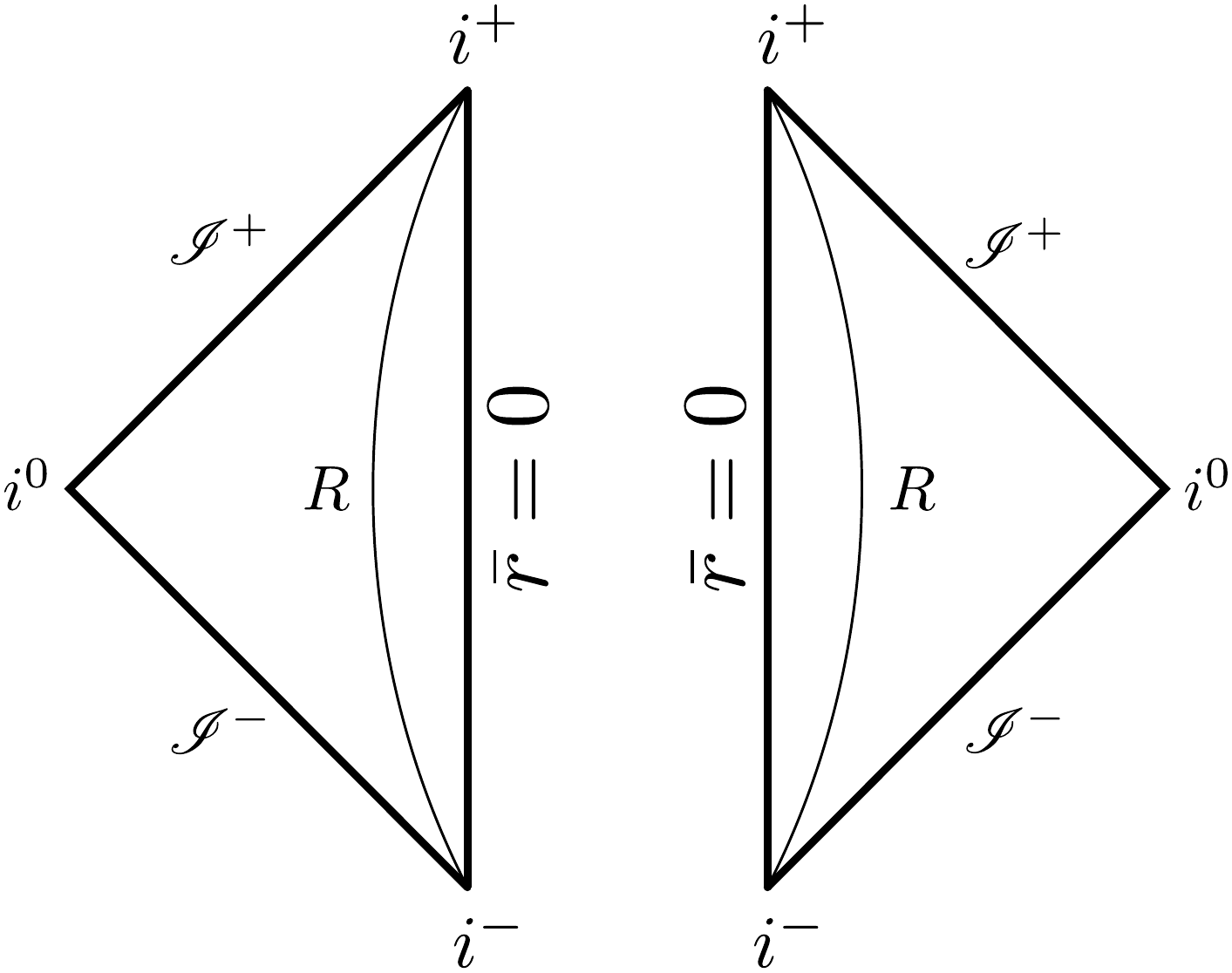}
\caption{\label{fig:Carter-Penrose_Friedmann_wormhole}
Carter-Penrose diagram of the Friedmann static open universe. The
vertical lines represent the poles of each hy\-per\-bo\-loid
branch, ${\bar
r}=0$, where $\bar r$ is the radial coordinate
related to the area defined by it, namely, ${\bar
r}=R\sinh\left(\frac{r}{R}\right)$. There is no hash
symbol \# here
because the two spacetimes
are disconnected,
no geodesic can pass from one spacetime to the other.
The
symbols $i^-$, $i^0$, and $i^+$ represent past timelike infinity,
spatial infinity, and future timelike infinity, respectively, and the
symbols $\mathscr{I}^-$ and $\mathscr{I}^+$, represent past and future
null infinity, respectively. The timelike line $R$ is drawn to call
attention that the Friedmann static open universe has a characteristic
intrinsic radius.
}
\end{figure}

The spatial and causal structure of
the spacetime
can also be presented.
Considering
a time slice $t={\rm constant}$ of the spacetime gives two copies of
the hyperbolic 3-space.  To see this, one makes an embedding. The
hyperbolic 3-space cannot be embedded in the 4-dimensional Euclidean
space, but it can be embedded to an open region of
the 4 dimensional Minkowski spacetime. Defining the Minkowski
coordinates $(w,x,y,z)$, where $w$ is a time coordinate and $(x,y,z)$
are spatial coordinates, as $w=R\cosh\frac{r}{R}$,
$x=R\sinh\frac{r}{R}\sin\theta\cos\phi$,
$y=R\sinh\frac{r}{R}\sin\theta\sin\phi$, and
$z=R\sinh\frac{r}{R}\cos\theta$, the line element of the embedded
surface is given by
$ds^2=-dw^2+dx^2+dy^2+dz^2$, and the surface verifies the equation 
$w^2-x^2-y^2-z^2=R^2$. So the surface, which represents
a time slice of the Friedmann
static universe, is in fact given by two copies of a 3-dimensional
hyperboloid.  To visualize the embedding, one makes a
$\theta=\frac{\pi}{2}$ slice, i.e., $z=0$ in the Minkowski
coordinates.  In Fig.~\ref{fig:time_slice_Friedmann_wormhole} we show
an embedding for the static hyperbolic open Friedmann universe.
Clearly there are two sheets, i.e., the universe has two branes, the
two copies of the Friedmann static universe.  Notice that we opted to
map the two asymptotic flat regions to open sets of the future and
past light cones to show both regions, although one should bear in
mind that, first, this has no physical relevance as both regions are
equivalent and, second, this has no relation with time reversal.
Moreover, admitting it might not be clear from the embedding diagram,
the vertices of the hyperbolas are identified as the same point hence,
the static hyperbolic Friedmann universe can also be seen as a model of
a failed wormhole where two asymptotic flat regions have a common
point with circumferential radius $R\sinh\frac{r}{R}=0$, i.e., $r=0$,
so that the wormhole's throat is a point, a zero measure set.
Since the two hyperboloid branes are independent there is indeed no
wormhole, it is a failed wormhole.
By making appropriate identifications each of the two open
infinite sheets turns into some closed 3-space, in which case
the volume of such a universe would be finite.  In
Fig.~\ref{fig:Carter-Penrose_Friedmann_wormhole} we show the causal
structure of the resulting spacetime in a Carter-Penrose diagram.  We
use the hash symbol \# to represent the
connected sum of the spacetime manifolds
in order to conserve the conformal structure in the Carter-Penrose
diagram of the total spacetime.  We see that it represents a universe
in which the spatial sections are two copies of a 3-hyperboloid.  The
Friedmann hyperbolic universe is a spacetime composed of time times
hyperbolic 3-space, actually two copies of it.  A geodesic, or a
free-falling particle, initially moving in the upper brane in the
direction of decreasing radial coordinate would reach $r=0$ and would
continue until it reaches infinity, without interacting with a mirror
geodesic, or a mirror free-falling particle, initially moving in the
lower brane in the direction of decreasing radial coordinate reaching
the same $r=0$ and continuing until it reaches the infinity of its own
brane.

The Friedmann static open universe is stable.  Indeed,
from Eq.~(\ref{stabEF}) one has that for $k=-1$,
$V''(R_{0})>0$.  This result
confirms the expectations.  For the static Friedmann  universe, a small
increase in the radius of the universe means less gravitational field
due to matter with negative energy density, so less repulsion,
and more cosmological tension field from $\Lambda$, so the universe
oscillates around the original radius in a stable situation.

This open static universe proposed by Friedmann came after a suggestion
by Fock, and was worked out by Friedmannbefore introducing, in
the same paper, the new expanding time-dependent hyperbolic solutions.
Friedmann's main motivation for presenting it was that the solution
represented the other side of Einstein's static universe, the two
solutions, Einstein's and Friedmann's, are indeed
complementary to each
other.  It is a much forgotten universe.  Since this static
solution has a negative energy density and a negative cosmological
constant, and violates all energy conditions, it seemed a physically
inadmissible strange universe that could hardly captured any
attention. This  prejudice against solutions that violate the
energy conditions came to an end when traversable wormholes, systems
that violate several energy conditions, jumped into the limelight. We
see here that the Friedmann static universe is a wormhole, albeit a
failed one.  Moreover, although the solution does not obey the energy
conditions, Friedmann's static universe is interestingly stable.  Thus,
Friedmann had a prescient foresight in contemplating working out
in detail the mathematics of this static solution.

We can now make a comparison between the Minkowski-Minkowski static
open universe, or traversable wormhole, and the Friedmann static
hyperbolic universe, or failed wormhole.  Although the two universes
are, of course, totally distinct solutions of the general relativistic
field equations, there are both differences and similarities between them,
although the smiliarites here are not so compelling.
In relation to the matter properties, the Minkowski-Minkowski open
universe, i.e., the traversable wormhole universe, is highly
nonuniform, it is vacuum everywhere except at a thin shell throat with
circumferential radius $R$, made of a perfect fluid with a negative
energy density $\sigma$ and a positive pressure $p$ to hold it static.
The Friedmann static open universe with characteristic radius $R$ is
uniform, permeated by a fluid with a repulsive negative energy density
$\rho$ and a negative, attractive, cosmological constant $\Lambda$ to
hold it static.  Thus, both spacetimes violate the energy condition,
they have negative energy densities and have some form of pressure,
positive tangential shell pressure in one case and negative
cosmological constant pressure in the other case, to hold them static.
In relation to the geometric and causal properties, one can compare
the figures drawn, namely, a $t=\text{constant}$ and
$\theta=\frac{\pi}{2}$ slice of the Minkowski-Minkowski open universe,
or traversable wormhole, and the corresponding Carter-Penrose diagram,
shown in Figs.~\ref{fig:time_slice_wormhole}
and~\ref{fig:Carter-Penrose_wormhole}, respectively, and a
$t=\text{constant}$ and $\theta=\frac{\pi}{2}$ slice of the Friedmann
open universe, or failed wormhole, and the corresponding
Carter-Penrose diagram shown in
Figs.~\ref{fig:time_slice_Friedmann_wormhole}
and~\ref{fig:Carter-Penrose_Friedmann_wormhole}. The comparison leads
to the conclusion that the two universes have some similarities.  Both
universes for large radii have two distinct open sheets, although the
circumferential radius in the Minkowski-Minkowski open universe is
finite not zero, and so composes a traversable wormhole, whereas in
the Friedmann static open universe the circumferential radius goes to
zero, and the wormhole fails to happen.  The bare geometrical
structure of the two universes is different, the Minkowski-Minkowski
open universe has geometry $\mathbb{R}\times\Sigma$, where $\Sigma$ is
a 3-space with nontrivial topology, and the Friedmann static open
universe has geometry $\mathbb{R}\times\mathbb{H}^{3}$, with negative
curvature in the two copies of the spatial sections.
The Minkowski-Minkowski open universe
could be thought of as being the
result of compressing, in a sense, the
evenly distributed matter of the Friedmann static universe into a thin
shell at some throat
radius leaving the rest of the spacetime empty.  
The
Minkowski-Minkowski open universe is stable, marginally, and the
Friedmann open universe is stable.
But here there are topological obstructions,
one universe is connected,
although not simply connected, the
other universe is disconected, it has two
separate branches, and so
one cannot pass from one universe to the other
without changing the topology.

\subsection{Einstein and Friedmann static universes:
One concept with two sides}

The Einstein and Friedmann static universes can be
seen as complementary to each other, i.e., they are two sides of the
same concept.  The concept, i.e., a collection of static
constant curvature
homogeneous  universes, yields on
one side of the concept a
closed universe, a bubble universe, and on the other
side of the concept
an open universe which is a failed wormhole.
The formalism presented in analyzing the two
universes is well suited to show this point.
From an algebraic point of view, one side of the concept is given by
$k=1$, the other side is given by
$k=-1$, where $k$ is a characteristic
that gives how space curves, positively in one case, negatively
in the other, respectively.
This algebraic side appears clearly in the evaluation of
the matter properties as displayed in
Eqs.~(\ref{eq:stability_sigma_value_generalEF})
and~(\ref{eq:stability_pressure_value_generalEF}).
More formally, to implement the idea of a closed
universe, i.e., a bubble universe,
and an open universe, i.e., a failed
wormhole, one uses the equations of general relativity.
For one universe one picks up $k=1$, a 3-dimensional
sphere. For the other universe
one picks up $k=-1$, a 3-dimensional
hyperboloyd.
From a matter point of view the two universes show a form of
complementarity, as for $k=1$ the matter obeys the energy conditions
while for $k=- 1$ the matter violates the energy conditions.
From a geometrical point of view, the two sides of the concept appear
when one picks up a manifold spacetime and at constant time
imposes a space with constant curvature.
One side is for positive curvature, a bubble universe,
the other side
for negative curvature, a failed wormhole.
Comparison of 
Fig.~\ref{fig:time_slice_Einstein_universe} with
Fig.~\ref{fig:time_slice_Friedmann_wormhole} for a spatial geometrical
representation of the bubble universe and the traversable wormhole,
respectively, displays some complementarity of the two spaces,
which can be further strengthened with the comparison of the spacetime
drawings in the form of Carter-Penrose diagrams, given in
Fig.~\ref{fig:Carter-Penrose_Einstein_universe} and in
Fig.~\ref{fig:Carter-Penrose_Friedmann_wormhole}, respectively.
From a stability point of view we have seen that
one universe is stable and the other unstable, showing thus some form
of complementarity.

So, that the two spacetimes, Einstein and Friedmann, demonstrate that
they can be seen as complementary to each other, i.e., they are two
sides of the same concept, is clear. It can be raised that the bubble
universe has matter that obeys the energy conditions, whereas the
failed wormhole has matter that does not obey. This is true, but again
there is no real problem.  In an early era of the Universe, when
quantum gravity dominates, there is no necessity of obeyance to the
energy conditions and the closed and open universes, created as bubble
universes and failed wormholes.  out of the spacetime foam they must
coexist together. Some kind of inflation would grow these objects to
macroscopic dimensions, making bubble universes and traversable
wormholes distinct, but connected, objects, some obeying the energy
conditions and being unstable, like the Eisntein universe, others
not obeying the energy conditions but being stable, like the
Friedmann static universe.

\section{Conclusions: Bubble universes and
traversable wormholes, two sides of one concept}
\label{Sec:Conclusions}

We have analyzed the possible universes that can be built from a
junction of two Minkowski spacetimes through a static, timelike thin
matter shell. Taking aside the trivial Minkowski flat universe with
no shell, there are two such universes.  One is a static closed
universe with a spherical thin shell with positive energy density and
negative pressure that joins two Minkowski balls, i.e., it
is the
Minkowski-Minkowski closed universe, a bubble universe.
The other universe is a
static open universe with a spherical thin shell with negative energy
density and positive pressure that joins two Minkowski asymptotic
sheets, it is the
Minkowski-Minkowski open universe, or traversable
wormhole. We have seen that they can be seen as complementary
to each other. More specifically, they are two sides
of one concept, the concept
being the collection
of nontrivial Minkowski-Minkovski spacetimes, with one side given by
$\frac12(\xi_{\rm i}-\xi_{\rm e})=1$,
the other side given by $\frac12(\xi_{\rm
i}-\xi_{\rm e})=-1$, where $\xi_{\rm i}$ and $\xi_{\rm e}$ are the
interior and exterior normals to the shell, respectively.

We have analyzed the possible universes that can be built from
static homegeneous pressurless matter with a cosmological constant.
There are two such universes.  One is the static closed Einstein
spherical universe. The other is the static open Friedmann hyperbolic
universe. We have seen that they can be seen as complementary to each
other, and, indeed, the idea of the construction of the static open
universe by Friedmann was to find the complement to the Einstein
universe.  More specifically, they are two sides of one concept, the
concept being the collection of constant curvature pressureless
universes, with one side given by positive curvature, $k=1$, the other
side given by negative curvature, $k=-1$.

We have also seen that the Minkowski-Minkowski closed universe, a
bubble universe, has resemblances with the static closed Einstein
universe with positive energy density, zero pressure, and positive
cosmological constant, and that the static open Minkowski-Minkowski
universe, a traversable wormhole, has resemblances with the static
open Friedmann universe with negative energy density, zero pressure,
and negative cosmological constant, which in turn is a failed
wormhole.  The Minkowski-Minkowski universes are both linearly stable,
marginally, and the Einstein and Friedmann static universes are
linearly unstable and stable, respectively.  One could think of the
Minkowski-Minkowski universes as being a limit of the homogeneous
universes when all the matter of the thin shell is spread evenly
throughout the universes, or vice versa, in which case
the homogeneous universes
being a limit of the Minkowski-Minkowski universes when all the matter
of the homogeneous universes is put somehow into thin shells.  For the
Einstein universe this would be possible classically, within general
relativity, as the two universes have the same topology, and so, for
constant universe radius, the Minkowski-Minkowski closed universe
could be the end point of the Einstein universe.  For the Friedmann
static universe this could not be realized within general relativity,
as the two universes have different topologies and so there is no way
of changing classically, and so continuously, from one into the other,
although quantum jumps of one to the other geometry might be
conceivably possible.

The existence of universes and wormholes within the Universe is a
tantalizing possibility allowed by the laws of physics.  Indeed, in a
early cosmic era when primordial scalar and gauge fields are dominant
and symmetry breaking phase transitions naturally arise, universes may
occur as bubbles within the Universe, and likewise, wormholes can
exist in the form of traversable shortcuts for distant parts of the
Universe or can even connect what would be distinct universes.  Bubble
universes and traversable wormholes are distinct
objects. Normally, bubble universes are found as dynamic solutions,
whereas, typically, traversable wormholes are studied as static
structures, but of course they can both be static or dynamic.  We have
analyzed two static cases, the two Minkowski-Minkowski spacetimes and
the two static homogeneous universes, and found that these spacetimes
demonstrate, in the way of example,
indeed two coupled examples that reinforce each other,
that bubble universes and
traversable wormholes can be seen as complementary to each other, i.e.,
they are two sides of some same concept. Dynamical cases where bubble
universes and traversable wormhole are complementary to each other can
also be found and studied. It is plausible that in a quantum gravity
scenario or in a scenario in which quantum gravity is weak but
nonnegligible, both bubble universes and traversable wormholes are
dynamically created alike, being as well two sides of the same concept.
In addition, using this duality, one can infer that, arbitrarily
advanced civilizations, with arbitrarily advanced technology to deal in
pratical terms with spacetime features, if they can build bubble
universes, they can also build traversable wormholes, and
conversely, if
they are apt to build traversable wormholes, as has been often
suggested, they should be apt to build bubble universes.

\section*{Acknowledgments}
JPSL acknowledges Funda\c{c}\~{a}o para a Ci\^{e}ncia e Tecnologia -
FCT, Portugal, for financial support through Project
No.~UIDB/00099/2020.  PL thanks Center for Mathematical Analysis,
Geometry and Dynamical Systems, Instituto Superior T\'{e}cnico and
Centro de Matem\'{a}tica, Universidade do Minho, where part of this
work has been performed, for the hospitality.

\end{document}